\documentclass[aps, prd, twocolumn, showpacs, superscriptaddress, groupedaddress]{revtex4} 
\usepackage{graphicx}	
\usepackage{amssymb}
\usepackage{dcolumn}
\usepackage[mathscr]{euscript}
\usepackage{amsfonts}
\usepackage{amsmath}
\usepackage{color}
\usepackage{subfigure}
\usepackage{epsfig}
\usepackage{subfigure, rotating, bm, array}
\usepackage[pagebackref=false, colorlinks=true]{hyperref}
\hypersetup{
linkcolor=blue,     
citecolor=blue,     
urlcolor=blue}      
\begin{document}

\title{Magnetic Penrose Process and Kerr Black Hole Mimickers}

\author{Divyesh P. Vithhani}
\email{divyeshviththani@gmail.com}
\affiliation{PDPIAS, Charusat University, Anand-388421 (Gujarat), India}

\author{Tapobroto Bhanja}
\email{tapobroto.bhanja@gmail.com}
\affiliation{PDPIAS, Charusat University, Anand-388421 (Gujarat), India}

\author{Vishva Patel}
\email{vishvapatelnature@gmail.com}
\affiliation{PDPIAS, Charusat University, Anand-388421 (Gujarat), India}

\author{Pankaj S. Joshi}
\email{psjcosmos@gmail.com}
\affiliation{International Centre for Space and Cosmology, Ahmedabad University, Ahmedabad, GUJ 380009, India}

\date{\today}

\begin{abstract}

The present study investigates the negative energy orbits and energy extraction efficiency using the magnetic Penrose process in various regular black hole geometries surrounded by electromagnetic fields. Utilizing numerical simulations, we analyze the efficiency of this process in Kerr and Simpson-Visser geometries, focusing on extremal black holes. Interestingly, our study demonstrates that the energy extraction efficiency remains indistinguishable between Kerr and Simpson-Visser geometries, regardless of the regularization parameter ($l$); this trend is consistent with previous studies of the Penrose process and superradiance. Additionally, we present results for the rotating Hayward black hole, showing that efficiency is influenced by spin and deviation parameters ($g$), as well as the induced magnetic field and charge of the compact object. Notably, we observe that energy extraction efficiency is highest in the rotating Hayward black hole compared to Kerr and Simpson-Visser geometries, particularly in scenarios where the magnetic field and charge are minimal. Our study highlights the significant role of spin, charge and magnetic field characteristics in maximizing energy extraction efficiency, particularly in the rotating Hayward black hole context.

\bigskip

$\boldsymbol{key words}$: Astrophysical compact objects, Energy extraction, Penrose process, Magnetic Penrose process 
\end{abstract}

\maketitle

\section{Introduction}

Recent discoveries of a significant number of high-energy astrophysical phenomena like gamma ray bursts, fast radio bursts, jet formations, etc., suggest that enormous large amounts of energy are continuously being produced and released, thus naturally invoking curiosity to study and understand them. The probable sources of these energy outbursts could be any active galactic nuclei, quasars, pulsars, system of binary stars or astrophysical compact objects. As strong gravitational fields generate energy and provide a means for converting one form of energy to another, it becomes essential and exciting to study the regions of strong gravitational fields to explain the astrophysical phenomena above. The critical questions that arise naturally are thus: (i) how such an enormous amount of energy being produced, (ii) would the nearby regions of any compact object produce such energy, and (iii) what are the limits and/or criteria of such energy extractions near compact objects or in regions of strong gravitational fields?\\ 

With this aim, several mechanisms have been studied in the literature, but mainly in the context of black holes. However, it is well-known that the end state of gravitational collapse could also result in a naked singularity \cite{joshi,goswami,mosani1,mosani2,mosani3,mosani4,Deshingkar:1998ge,Jhingan:2014gpa,Joshi:2011zm,Joshi:2023ugm}. The event horizon is the most distinct property distinguishing black holes from a naked singularity. Quite a significant amount of work has been done in recent years to study the naked singularity and a black hole with respect to accretion properties, shadows and tidal forces \cite{Stuchlik:2024xrw,Patel:2022vlu,Solanki:2021mkt,Joshi:2013dva,Tahelyani:2022uxw,Kovacs:2010xm,Guo:2020tgv,Chowdhury:2011aa,Paul:2020ufc,Pugliese:2014ela,Bambhaniya:2019pbr,Dey:2019fpv,Dey:2020bgo,Dey:2020haf,Kaur:2021wgy,Page:1974he,Liu:2020vkh,Bambhaniya:2021ugr,Pugliese:2024bhh,Rahaman:2021kge,Harko:2008vy,Harko:2009xf,Lattimer:1976kbf,Shaikh:2019jfr,Vachher:2024ldc,Shaikh:2018oul,Kumar:2022fqo,Vrba:2023byq,Eva2,Narzilloev:2021jtg,AlZahrani:2014dfi,Narzilloev:2021ygl,Shaymatov:2021qvt,Bam2020,Pugliese:2013xfa,AlZahrani:2022fas,Bambhaniya:2022xbz,Gralla:2019xty,Shaikh:2022ivr,Ghosh:2020ece,Vagnozzi:2019apd,Chen:2022nbb,atamurotov_2015,abdujabbarov_2015b,Li:2021,Hu:2020usx,Saurabh:2020zqg,Pugliese:2022oes,Vagnozzi:2022moj,Saurabh:2022jjv}. \\

The initial attempt to understand the phenomenon of energy extraction was made by Penrose and Floyd, who studied the mechanism of rotational energy extraction from a Kerr black hole \cite{Penrose:1971uk}. It was followed by the study of the Penrose process in quasars, the presence of an external source-less magnetic field, which came to be known as the magnetic Penrose process. In both mechanisms, it has been shown that, theoretically, energy can be extracted from the black holes due to the presence of negative energy orbits and the ergoregion. In the Penrose process, as a particle enters the ergoregion or the effective ergoregion, it splits up into two, one entering the black hole and has a $negative$ energy (as compared to an observer at infinity), thereby reducing the total energy of the black hole. At the same time, the other part escapes to infinity with an energy greater than that of the incident particle. In this context, it is worth mentioning that Christodoulou \cite{Christodoulou1} has argued that energy can only be extracted from a black hole.  The other well-studied mechanisms of energy extraction around a compact object are the collisional Penrose process, radiative Penrose process, rotational energy extraction from accretion discs,  Blandford-Znajek process (BZ), Blandford Payne process (BP), Banados-Silk-West (BSW) process \cite{Patel:2022jbk,Franzin:2022iai,Zaslavskii:2020kpv,Pugliese:2021ivl,Acharya:2023vlv,Zaslavskii:2022nbm,Patel:2023efv,Patil:2011yb,Banados:2009pr,Feiteira:2024awb,Patil:2011aa,Patil:2011uf,Patil:2011ya,Patil:2011aw,Chakraborty:2024aug,Zaslavskii:2024zgh}. In the collisional Penrose process, one considers the collision of particles near a black hole. Blandford\,-\,Znajek and Blandford\,-\,Payne processes consider the interplay of magneto-hydrodynamics and a rotational compact object as a source. It is worth mentioning here that the acceleration (or retardation) of the colliding particles is also considered in the BSW effect. \\

As discussed earlier, in the study of energy extraction from and around compact objects,  most of the works done till now harbour spacetime singularities (both black holes and naked singularities have been studied) where the quantities, like mass density and curvature of space-time, etc.,  have a diverging nature. The problem of singularities can, in principle, be resolved by (i) considering quantum gravity, which would remove spacetime singularities when length scales comparable to or less than the Planck length are being considered, or (ii) by studying the regular spacetime geometries, which by definition do not harbour a central singularity, thereby not bringing in any limitations to general relativity and evading incompleteness of the spacetime geometry. Thus, these spacetime geometries make them an interesting prospect for study in astrophysics. In this direction,  the first regular black hole solution was proposed by Bardeen \cite{Bardeen1},
which has a regular centre and behaves like a Schwarzschild black hole asymptotically so that the incompleteness of the spacetime is removed. Notably, it has been shown that one can derive Bardeen spacetime if one considers non-linear electrodynamics \cite{Bardeen3}.  In the case of Bardeen spacetime, magnetic monopole originating from the consideration of non-linear electrodynamics have also been studied \cite{Bardeen2}. Some other well-known regular spacetimes are the Hayward spacetime and the Simpson-Visser spacetime \cite{Hayward,SimpsonV}. The static version of these spacetimes has been extensively studied from an astrophysical point of view in the realms of shadows, particle trajectories, tidal forces, and quasi-normal oscillations \cite{Lobo, Simpson1, Adolfo, Balart,  Kumar, sushantghosh, Neves, Lemos, Zaslavskii, Fan, Dymnikova}, etc.  \\

In our present work, we focus on the rotating versions of the above-mentioned spacetimes viz.,  the rotating Simpson-Visser, the rotating  Regular Hayward Spacetime (Simpson-Visser and Hayward spacetime are regular black hole spacetimes), and the rotating Janis-Newman-Winicour naked singularity to study the process of energy extraction under two conditions: (i) in the presence of an external magnetic field and (ii) in its absence. The motivation for considering an external magnetic field comes from the fact that, in principle, energy extraction in the presence of a magnetic field should be significantly different from its absence around a rotating compact object, while the motivation for studying both rotating black hole spacetimes and rotating naked singularity spacetime is to study the effect of the absence or presence of an event horizon on energy extraction efficiency. Thus, it becomes an important and interesting endeavour to study the phenomenon of energy extraction around the above-mentioned rotating compact objects in the presence and absence of an external magnetic field. In this context, it is interesting to see the behaviour of JMN-1 naked singularity when considering the external magnetic field \cite{KaunteyJoshi}.  \\

This article is organized as follows: Sec. (\ref{Sec_2RSV}) describes the preliminaries about the rotating Simpson-Visser spacetime. Depending on the value of the spin parameter ($a$), the ADM mass ($M$), and the regularization parameter ($l$),  we discuss the six possibilities, categorizing them into either regular black hole or wormhole spacetimes. In Sec. (\ref{Sec_3rjnw}), we discuss the rotating Janis-Newman-Winicour naked singularity and our Sec. (\ref{Sec-4rh}) deals with the rotating Hayward Spacetime. Sec. (\ref{sec_magnetic}) is concerned with the mechanism of introduction of an external magnetic field around the (considered) compact object. Our Sec. (\ref{sec_neo}) is dedicated to negative energy orbits and the energy extraction process from near the compact objects. Finally, we conclude our present work with discussions regarding this study in Sec. (\ref{Sec_discussion}). Throughout the paper, we have used the geometrized units: $8\, \pi\, G = c = 1$. \\

\begin{figure*}[]
\centering
\subfigure[$M\,=\,1, Q\,=\,0.1, l\,=\,0.4 \,$\,(RBH-2).]
{\includegraphics[width=5.7cm]{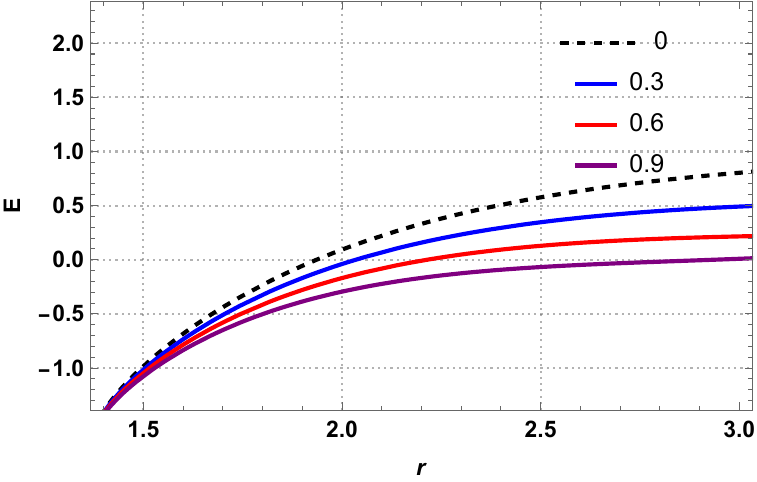}\label{fig-011}}
\subfigure[$M\,=\,1, Q\,=\,0.1, l\,=\,0.8$\,(eRBH).]
{\includegraphics[width=5.7cm]{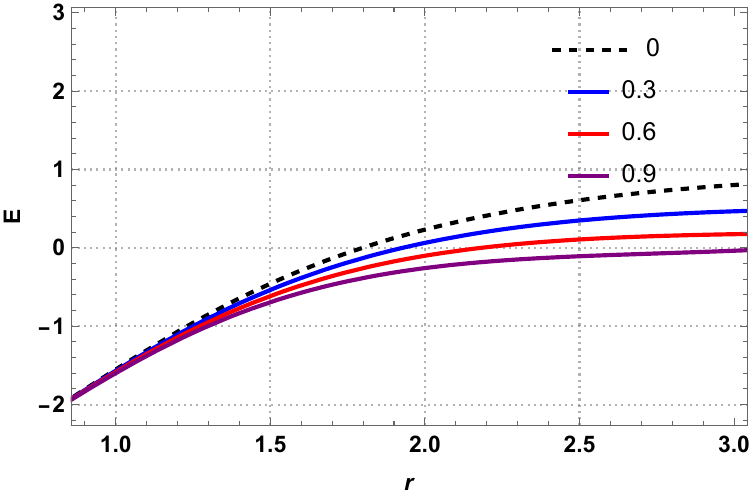}\label{fig-012}}
\subfigure[$M\,=\,1, Q\,=\,0.1, l\,=\,0.8$\,(RBH-1).]
 {\includegraphics[width=5.7cm]{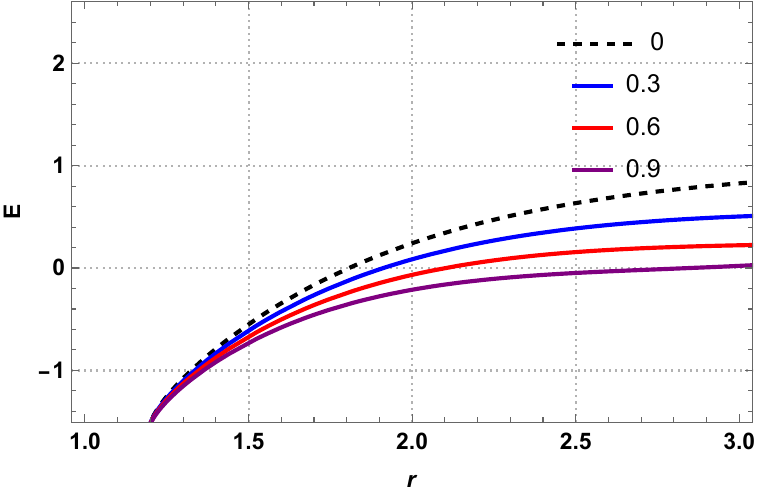}\label{fig-013}}
 \subfigure[$M\,=\,1, Q\,=\,0.1, l\,=\,0.5641$\,(nRBH).]
{\includegraphics[width=5.7cm]{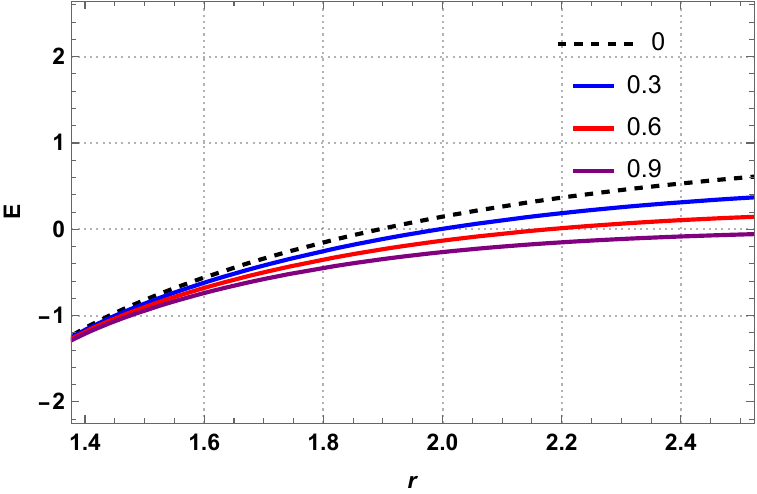}\label{fig-014}}
\subfigure[$M\,=\,1, Q\,=\,0.1, l\,=\,1.4358$\,(nWoH).]
{\includegraphics[width=5.7cm]{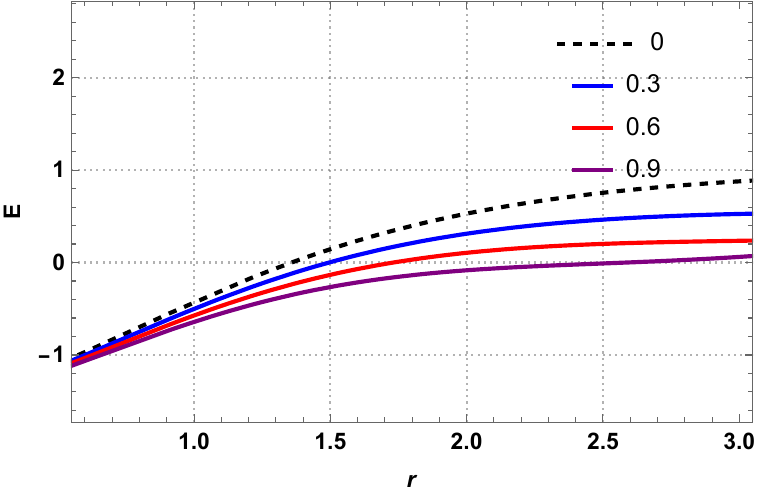}\label{fig-015}}
\subfigure[$M\,=\,1, Q\,=\,0.1, l\,=\,1.8$\,(WoH).]
{\includegraphics[width=5.7cm]{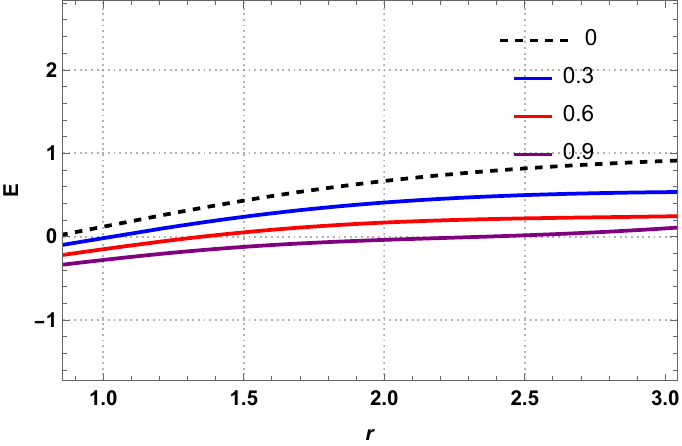}\label{fig-016}}
\subfigure[$M\,=\,1, Q\,=\,0.9, l\,=\,0.4$\,(RBH-2).]
 {\includegraphics[width=5.7cm]{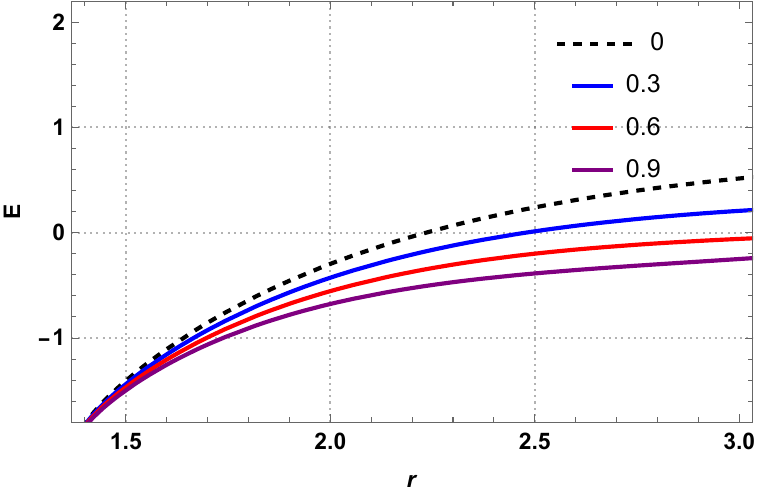}\label{fig-017}}
\subfigure[$M\,=\,1, Q\,=\,0.9, l\,=\,0.8$\,(eRBH).]
{\includegraphics[width=5.7cm]{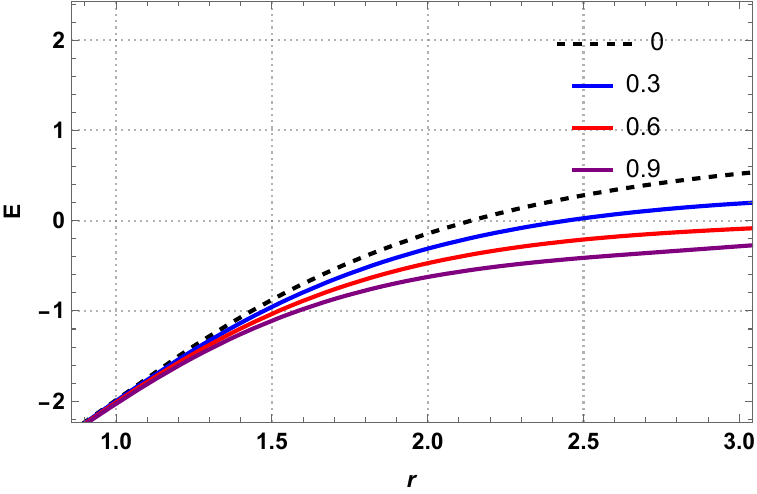}\label{fig-018}}
\subfigure[$M\,=\,1, Q\,=\,0.9, l\,=\,0.8$\,(RBH-1).]
{\includegraphics[width=5.7cm]{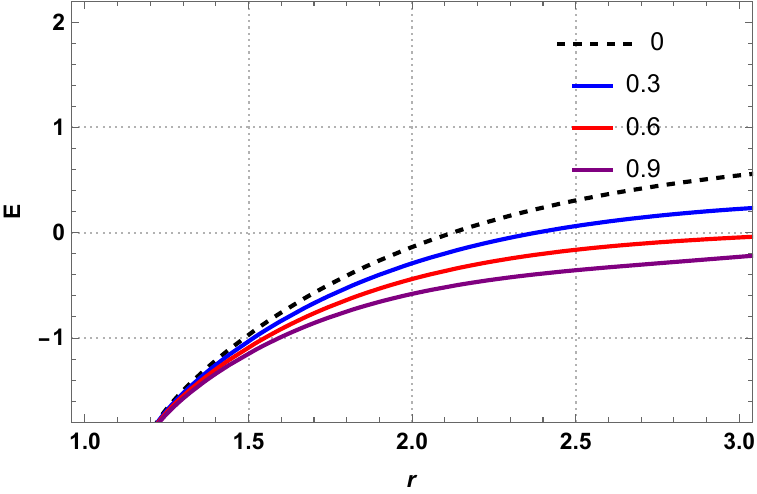}\label{fig-019}}
\subfigure[$M\,=\,1, Q\,=\,0.9, l\,=\,0.5461$\,(nRBH).]
 {\includegraphics[width=5.7cm]{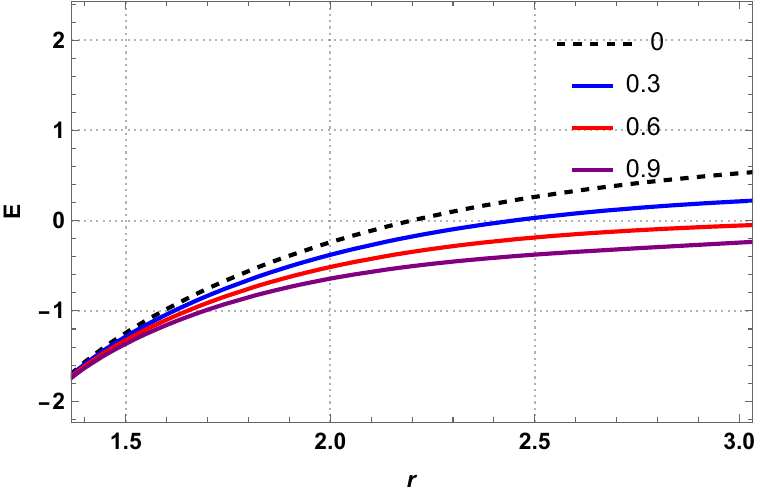}\label{fig-110}}
 \subfigure[$M\,=\,1, Q\,=\,0.9, l\,=\,1.4358$\,(nWoH).]
{\includegraphics[width=5.7cm]{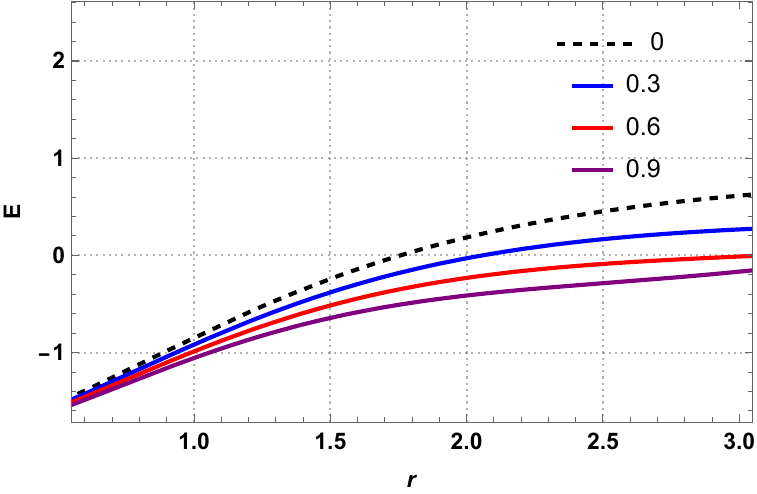}\label{fig-111}}
\subfigure[$M\,=\,1, Q\,=\,0.9, l\,=\,1.8$\,(WoH).]
{\includegraphics[width=5.7cm]{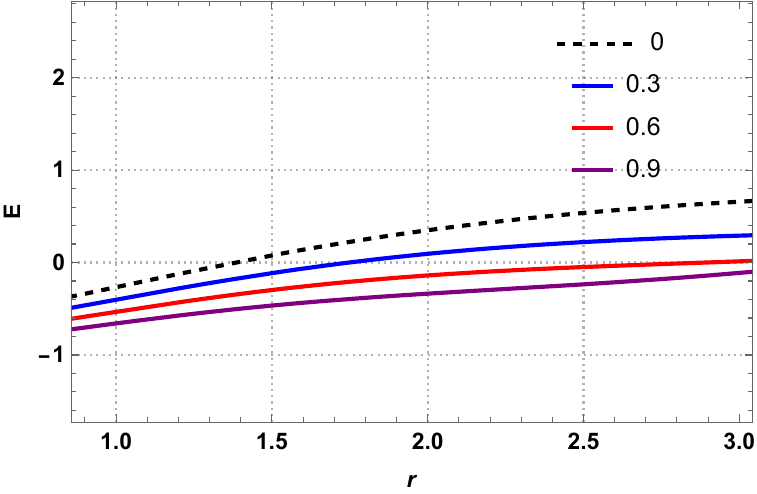}\label{fig-112}}
\subfigure[$M\,=\,1, Q\,=\,0, l\,=\,0.8, a\,=\,0.9$.]
{\includegraphics[width=5.7cm]{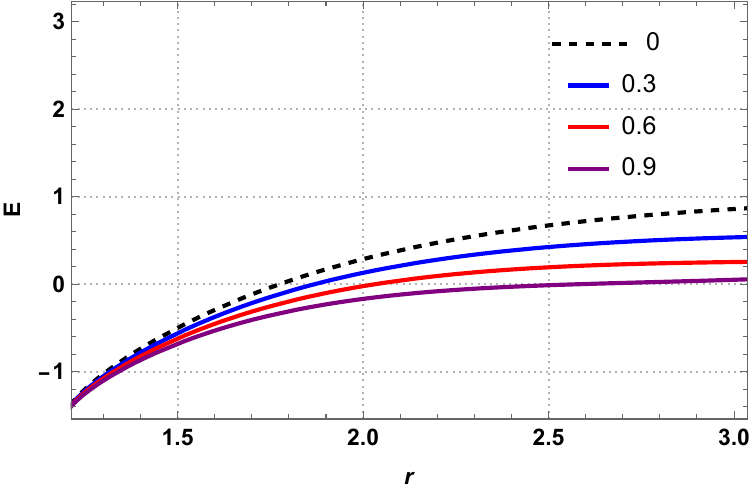}\label{fig-113}}
\subfigure[$M\,=\,1, Q\,=\,0, l\,=\,1.6, a\,=0\,.9$.]
{\includegraphics[width=5.7cm]{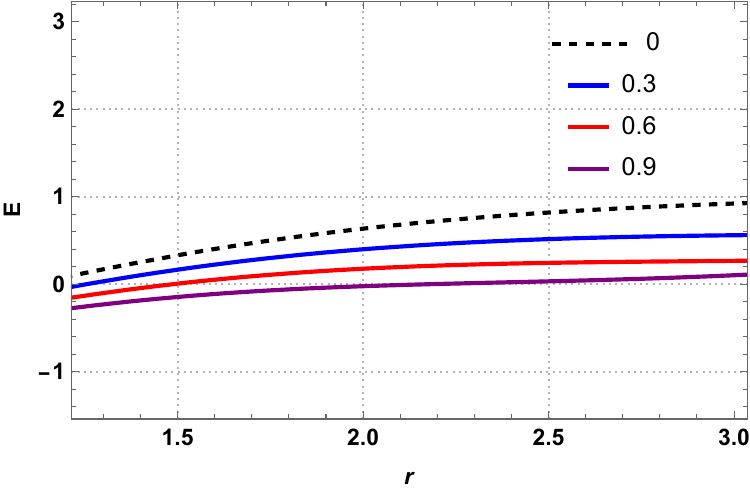}\label{fig-114}}
\subfigure[$M\,=\,1, Q\,=\,0, l\,=\,2.4, a\,=\,0.9$.]
{\includegraphics[width=5.7cm]{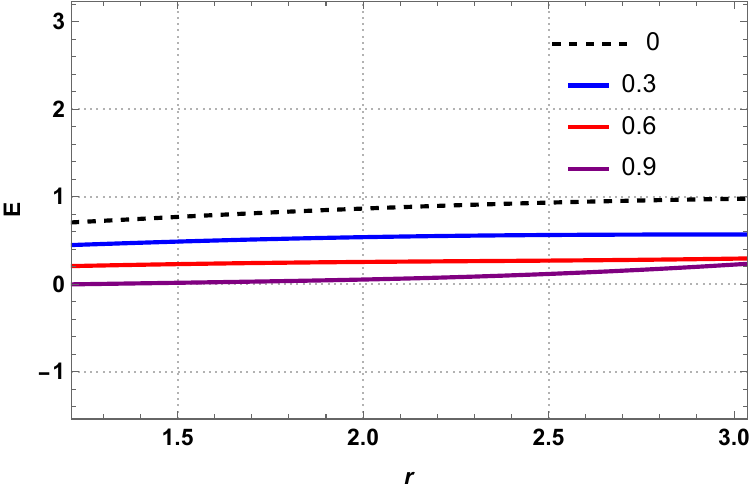}\label{fig-115}}
\caption{The above figures show the variation in total energy as the radial distance changes in the rotating Simpson-Visser geometry for values of different parameters. The values displayed on the right-hand side of the plots indicate the magnetic field strength ($B$) around the compact object. It is assumed that the angular momentum ($L$) of a charged particle is -5 and the spin ($a$) of a compact object is 0.9, except for eRBH case $a$, which is considered the same as mass.
}\label{fig-1} \end{figure*}

\begin{figure*}[]
\centering
\subfigure[$M\,=\,1, Q\,=\,0.1, g\,=\,0.1, a\,=\,0.8$.]
{\includegraphics[width=5.7cm]{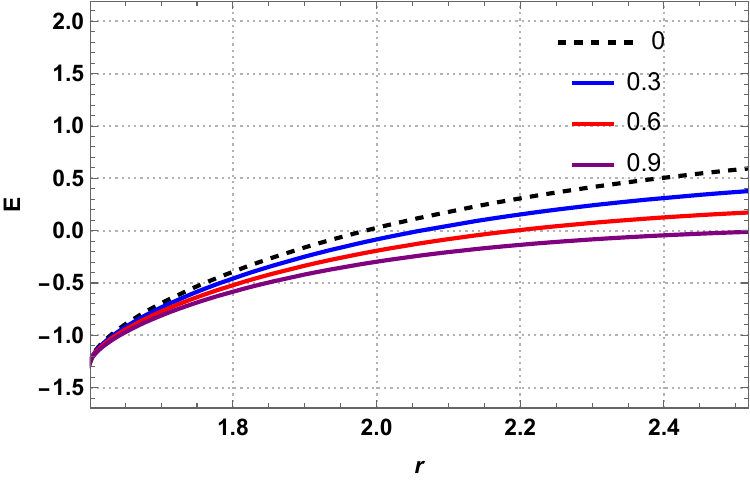}\label{fig-021}}
\subfigure[$M\,=\,1, Q\,=\,0.1, g\,=\,0.2, a\,=\,0.99$.]
{\includegraphics[width=5.7cm]{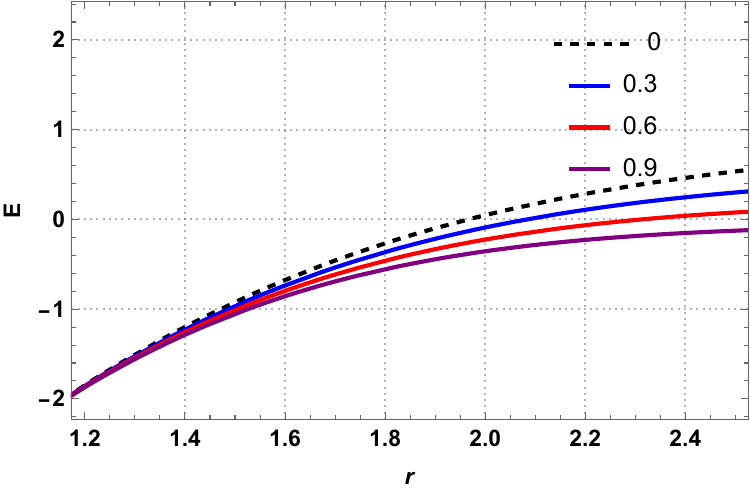}\label{fig-022}}
\subfigure[$M\,=\,1, Q\,=\,0.1, g\,=\,0.4, a\,=\,0.9$.]
 {\includegraphics[width=5.7cm]{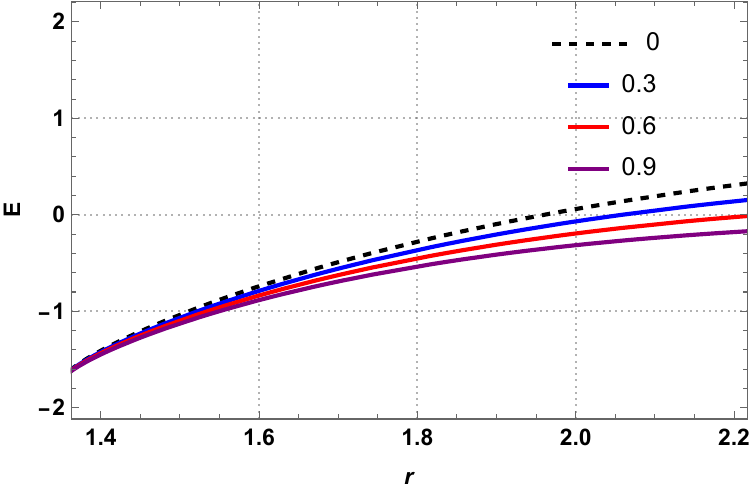}\label{fig-023}}
 \subfigure[$M\,=\,1, Q\,=\,0.1, g\,=\,0.6, a\,=\,0.8$.]
{\includegraphics[width=5.7cm]{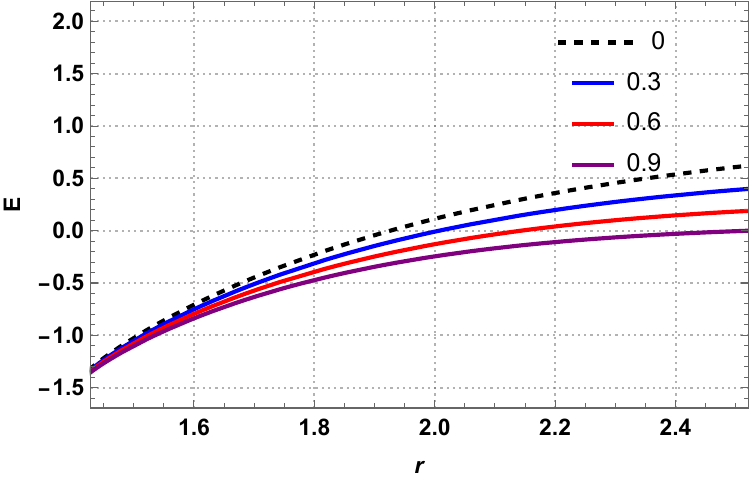}\label{fig-024}}
\subfigure[$M\,=\,1, Q\,=\,0.1, g\,=\,0.7, a\,=\,0.7$.]
{\includegraphics[width=5.7cm]{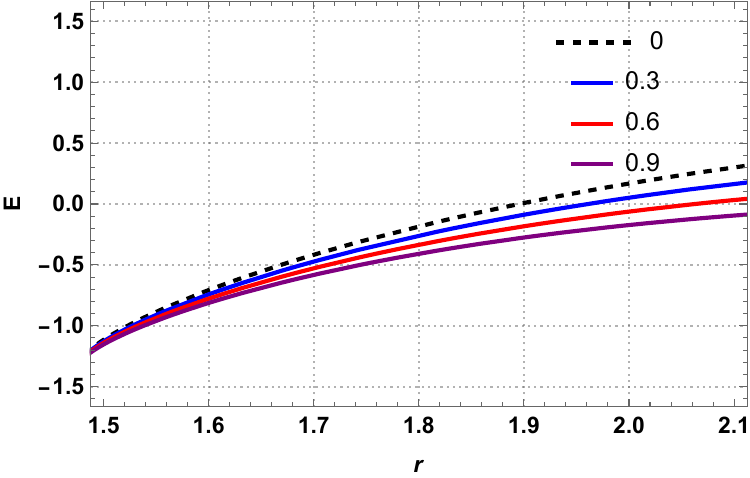}\label{fig-025}}
\subfigure[$M\,=\,1, Q\,=\,0.1, g\,=\,0.8, a\,=\,0.3$.]
 {\includegraphics[width=5.7cm]{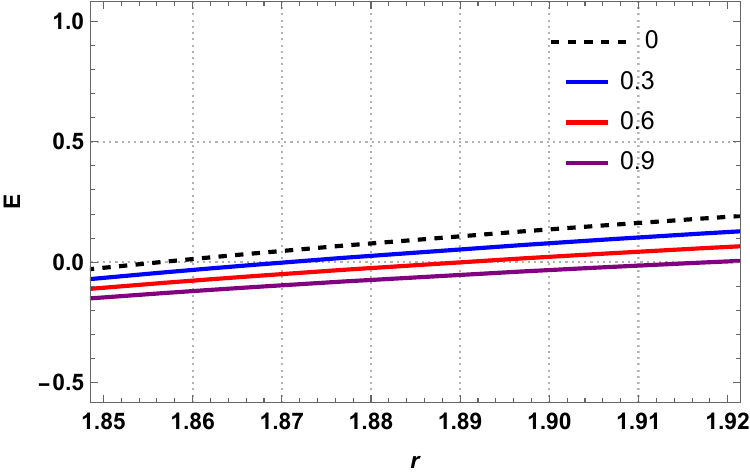}\label{fig-026}}
\subfigure[$M\,=\,1, Q\,=\,0.9, g\,=\,0.1, a\,=\,0.8$ .]
{\includegraphics[width=5.7cm]{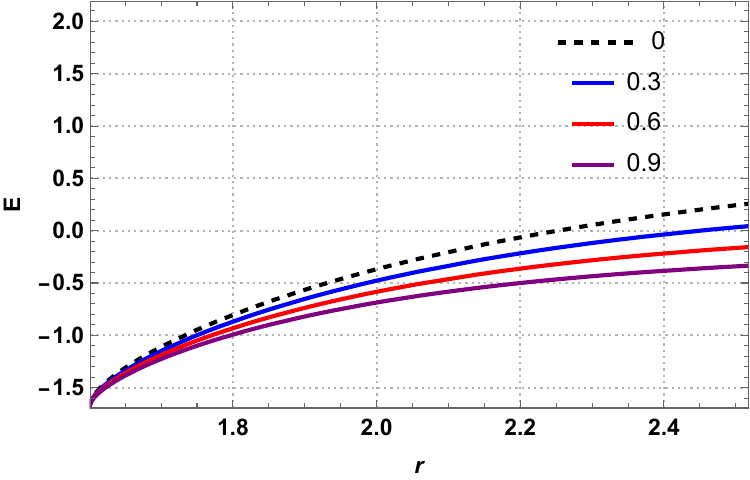}\label{fig-027}}
\subfigure[$M\,=\,1, Q\,=\,0.9, g\,=\,0.2, a\,=\,0.99$.]
{\includegraphics[width=5.7cm]{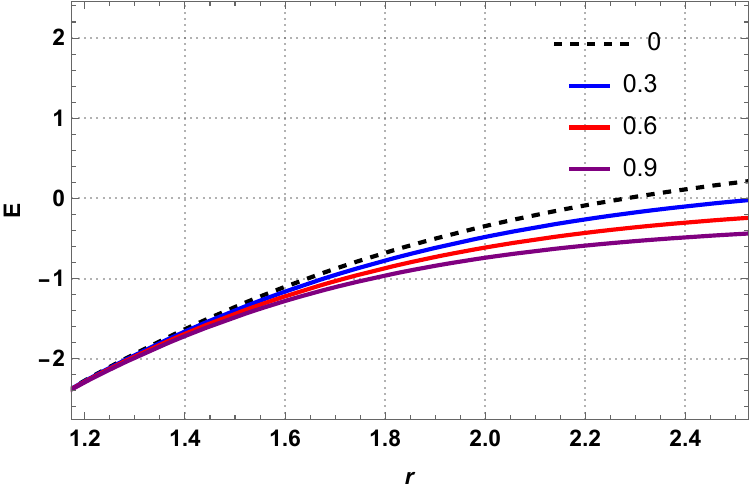}\label{fig-028}}
\subfigure[$M\,=\,1, Q\,=\,0.9, g\,=\,0.4, a\,=\,0.9$.]
 {\includegraphics[width=5.7cm]{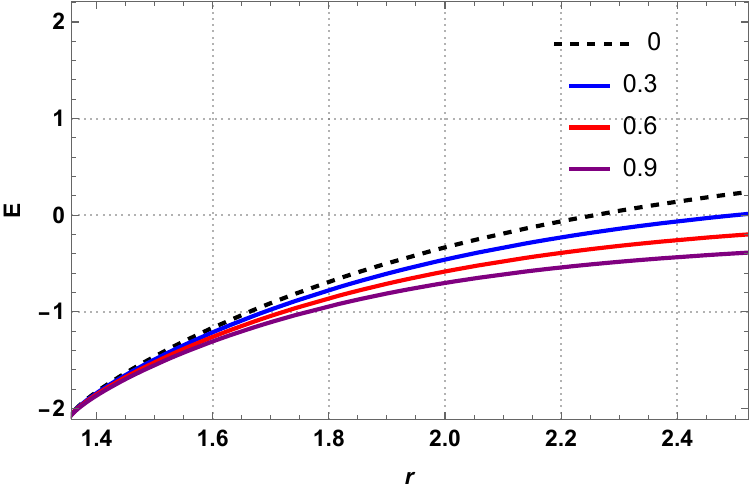}\label{fig-029}}
 \subfigure[$M\,=\,1, Q\,=\,0.9, g\,=\,0.6, a\,=\,0.8$.]
 {\includegraphics[width=5.7cm]{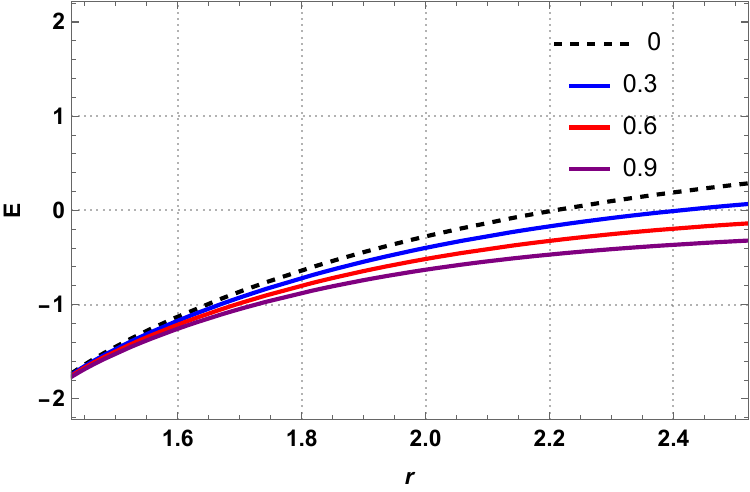}\label{fig-210}}
 \subfigure[$M\,=\,1, Q\,=\,0.9, g\,=\,0.7, a\,=\,0.7$.]
 {\includegraphics[width=5.7cm]{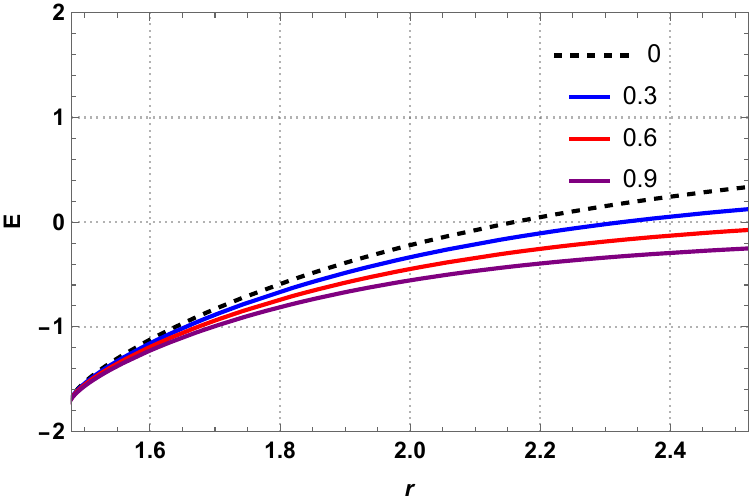}\label{fig-211}}
 \subfigure[$M\,=\,1, Q\,=\,0.9, g\,=\,0.8, a\,=\,0.3$.]
 {\includegraphics[width=5.7cm]{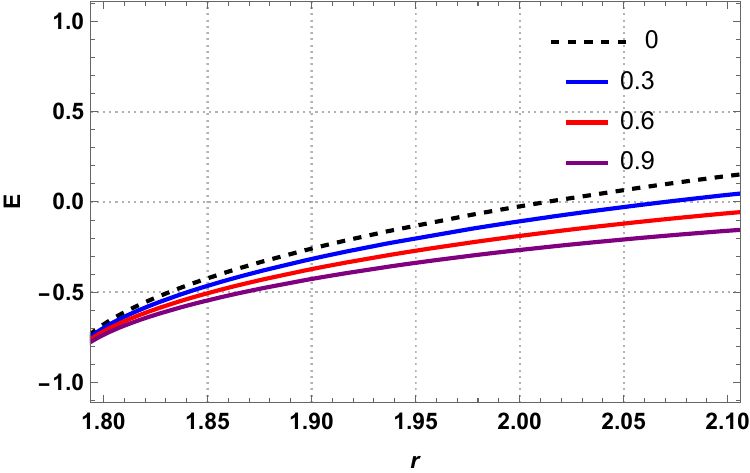}\label{fig-212}}
 \subfigure[$M\,=\,1, Q\,=\,0, g\,=\,0.8, a\,=\,0.9$.]
 {\includegraphics[width=5.7cm]{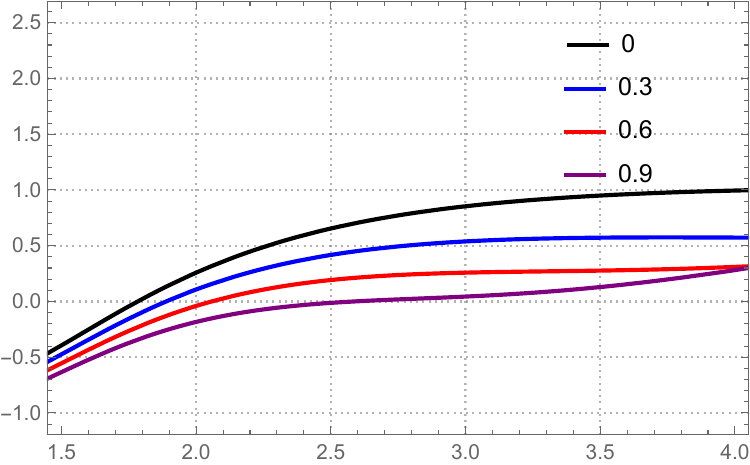}\label{fig-213}}
 \subfigure[$M\,=\,1, Q\,=\,0.5, g\,=\,0.8, a\,=\,0.9$.]
 {\includegraphics[width=5.7cm]{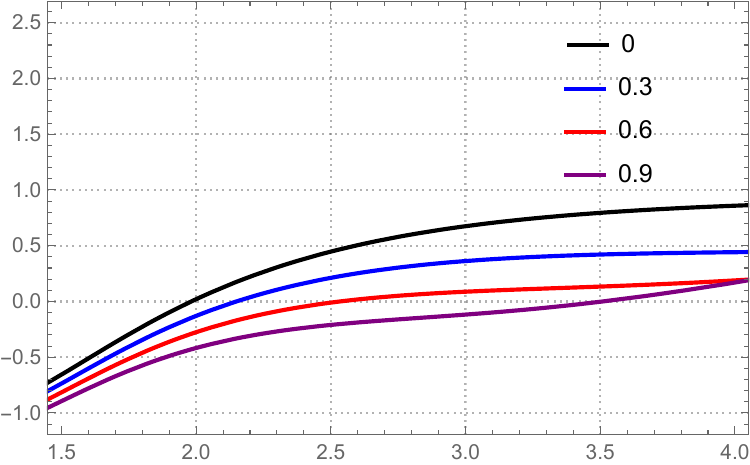}}\label{fig-214}
  \subfigure[$M\,=\,1, Q\,=\,0.9, g\,=\,0.8, a\,=\,0.9$.]
 {\includegraphics[width=5.7cm]{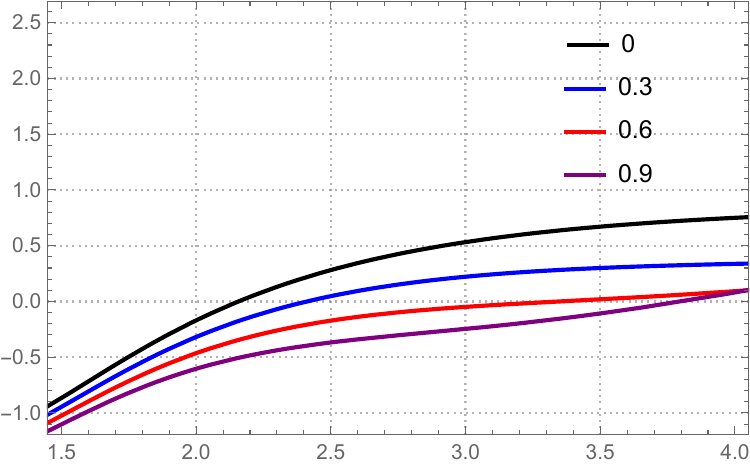}}\label{fig-215 }
\caption{The figures above illustrate how the total energy changes with radial distance in the regular rotating Hayward black hole geometry for different parameter values. The numerical values on the plots' right side represent the magnetic field strength ($B$) surrounding the compact object. It is assumed that the charged particle's angular momentum ($L$) is -5.}\label{fig-2}
\end{figure*}

\section{ROTATING SIMPSON-VISSER GEOMETRY}
\label{Sec_2RSV}

The static Simpson-Visser geometry is a form of a regularised Schwarzschild spacetime. Using the Newman–Janis algorithm without complexification, we can obtain a rotating version of the Simpson-Visser black hole \cite{Mazza:2021rgq}. The line element of rotating Simpson-Visser spacetime can be written as,

\begin{multline}
     ds^2 =-\left(1-\frac{ 2 f_{S}}{\rho_{S}^2} \right) \, dt^2 - \frac{ 4 a f_{S} \sin^2 \theta}{\rho_{S}^2} \, dt\,d\phi \\
     + \frac{\Sigma_S \sin^2 \theta}{\rho_{S}^2} \,d\phi^2  + \frac{\rho_{S}^2}{\Delta_{S}} \, dr^2 + \rho_{S}^2 \, d\theta^2,
    \label{SVmetric}
\end{multline} 

\begin{eqnarray}
    f_S = M \sqrt{ r^2 + l^2 },
    \label{fs} \nonumber \\
    \rho^2_s = r^2 + l^2 +a^2 \cos^2 \theta,
    \label{rhos}\nonumber \\
 \Delta_s =  r^2 + l^2 + a^2 -2 M \sqrt { r^2 + l^2 },
    \label{deltas}\nonumber \\
    \Sigma_s = ( r^2 + l^2 + a^2 )^2 -\Delta_s a^2 \sin^2 \theta .
\end{eqnarray}
   
Where $M$, $a$, and $l$ are the ADM mass, spin, and regularisation parameters. $l$ is a non-negative constant with the dimension of length. This spacetime represents various classes of solutions (black hole, regular black hole and wormhole) depending on the value of $a$ and $l$. One can locate the event-horizon by equating $\Delta_{s}\,=\,0$ ($g_{rr}\,\rightarrow \, \infty$),

\begin{equation}
    r_\pm = r \pm [(M \pm \sqrt{M^2 - a^2})^2 -l^2]^\frac{1}{2}.
    \label{reh}
\end{equation}

Also, consider

\begin{equation}
    \Xi = M \pm \sqrt{M^2 - a^2}.
\end{equation}

Here, we have categorized possible geometry cases depending on the value of $l$ and $a$. 

\begin{enumerate}
    \item \textbf{WoH} is a traversable wormhole ($l> \Xi_+$, $a<M$ or $a>M$). This geometry has a two-way wormhole with a timelike throat at $r=0$. The Penrose diagram for this geometry is similar to that of Minkowski's; one can distinguish regions of geometry as $r>0$ as ``our universe" and $r<0$ as ``other universe".
    
    \item  \textbf{nWoH} is a null wormhole; ($l=\Xi_+$ and $a<M$) in which an extremal event horizon exists and is a one-way wormhole with a null throat at $r=0$.

    \item \textbf{RBH-I} is a regular black hole; ( $\Xi_- < l < \Xi_+$ and $a < M$ ) and a geometry that contains a wormhole with a regular spacelike throat which is enveloped by an event horizon.

    \item \textbf{RBH-II}  is a regular black hole with two event horizons on each side ($l<\Xi_-$ and $a<M$). The geometry of this system can be described by a Carter Penrose diagram. This geometry consists of two Kerr-like patches glued at the throats. The throat is time-like, meaning it is transverse in both directions. However, the horizons are event horizons. As a result, an observer who crosses the throat twice cannot return to the asymptotically flat region.

    \item \textbf{eRBH} is the extremal version of RBH-II. ($l<\Xi_-=\Xi_+$ and $a=M$), the nature of the throat is time-like, and the two event horizons of this regular black hole coincide.

    \item \textbf{nRBH} is null RBH-I ($l=\Xi_-$ and $a<M$). This geometry is a regular black hole with one event horizon on each side and containing a null-like throat.
\end{enumerate}


\section{ROTATING JANIS-NEWMAN-WINICOUR NAKED SINGULARITY}
\label{Sec_3rjnw}

Janis-Newman-Winicour spacetime is an extended geometry of Schwarzschild spacetime with a mass-less scalar field. Due to a mass-less scalar field, rotating Janis-Newman-Winicour spacetime contains a rotating naked singularity instead of a black hole. Using the Newman–Janis algorithm without complexification, one can obtain a rotating version of Janis-Newman-Winicour spacetime similar to the previous spacetime\,\cite{Solanki:2021mkt}. The line element of rotating Janis-Newman-Winicour spacetime is,

\begin{multline}
   ds^2=-\left(1-\frac{ 2 f_{J}}{\rho_{J}^2} \right) dt^2 - \frac{ 4 a f_{J} \sin^2 \theta}{\rho_{J}^2}dt d\phi + \frac{\Sigma_J \sin^2 \theta}{\rho_{J}^2} d\phi^2 \\ + \frac{\rho_{J}^2}{\Delta_{J}} dr^2 + \rho_{J}^2 d\theta^2,
    \label{JNWSmetric}
\end{multline}
where,
\begin{eqnarray}
    &&\nu = \frac{2 M} {b},
    \label{nu} \quad  b = 2 \sqrt{ M^2 + q^2_s },
    \label{b}\nonumber\\
    &&  f_J = \frac{1}{2} \left ( \left ( 1 - \frac{ 2 M } { \nu r } \right) ^{-\nu} - 1 
 \right) \left ( 1 - \frac{ 2 M } { \nu r } \right),
 \label{fJ}\nonumber\\
&& \rho_J^2 = a^2 \cos^2 \theta + r^2 \left( 1 - \frac{ 2 M }{ \nu r } \right)^{ 1 - \nu },
    \label{rhoJ}\nonumber\\
 &&  \Sigma_J = ( a^2 \sin^2 \theta + \rho_J^2 )^2 - a^2 \sin^2 \theta \Delta_J,
    \label{sigmaJ} \nonumber\\
 &&\Delta_J = a 2 -\frac{2 M r}{\nu} + r^2,
\label{deltaJ}   
\end{eqnarray}

and $M$, $a$ and $q_{s}$ are the ADM mass, spin parameter and scalar field charge, respectively. The value of $\nu$ must be non-zero and within the $ 0 < \nu < 1 $ range. For $\nu=1$ and $a=0$, it corresponds to a Schwarzschild black hole; for $\nu=1$, it corresponds to a Kerr black hole. 


\section{REGULAR ROTATING HAYWARD BLACK HOLE}
\label{Sec-4rh}
The spherically symmetric static spacetime geometry of the regular rotating Hayward black hole can be described by the metric in Boyer-Lindquist coordinates as \cite{Khan:2021wzm}, 

\begin{multline}
     ds^2 = -(1-\frac{2f_{R}}{\rho_R^2})dt^2 -\frac{4a f_{R} \sin^2\theta}{\rho_R^2} dt d\phi \\ + \sin^2 \theta \left( a^2 +r^2 + \frac{2a^2r \Lambda \sin^2 \theta}{\rho_R^2 } \right)d\phi^2 + \frac{\rho_R^2}{\Delta_R}dr^2 + \rho^2_Rd\theta,
     \label{RHB}
\end{multline}
    where,
\begin{eqnarray}
&&   f_R = r\Lambda,
    \label{Delta M}   \,\, \Delta_R = r^2 - 2r \Lambda + a^2
    \label{Delta M1}, \nonumber\\
  &&  \rho_R = r^2 + a^2 \cos^2 \theta,  \nonumber\\
 && \Lambda = M \frac{r^{\kappa + 3} \rho^{-\kappa}}{r^{\kappa + 3} \rho^{-\kappa} + g^3 r^\psi \rho^{-\psi}}.
\end{eqnarray}
  
The parameters $a$ and $M$ represent the spin and mass of the black hole, respectively, while $g$ represents the deviation parameter, which provides a deviation from the standard Kerr black hole. Both $\kappa$ and $\psi$ are positive real numbers. When considering the equatorial
plane, the mass function takes the given form,

\begin{equation}
    \Lambda = M\frac{r^3}{r^3 + g^3}.
\end{equation}

The above equation does not depend on $\kappa$, $\psi$, and $\theta$. This spacetime geometry consists of an event horizon depending on the value of $a$ and $g$, as discussed in \cite{Khan:2021wzm}. For $g=0$, this metric reduces to the Kerr metric, and for $a=g=0$, it reduces to the Schwarzschild black hole.


\section{Magnetized rotating compact object}
\label{sec_magnetic}

As we have considered an asymptotic uniform magnetic field around the compact object, and the intensity of the magnetic field is defined by the parameter $B$. The direction of the magnetic field line is perpendicular to the equatorial plane of the spacetime geometry. We have also considered the induced charge of the compact objects (denoted by $Q$). This can be described using the components of electromagnetic four vector  $A_\mu$ with two non-zero components\,\cite{Stuchlik:2019dlx} as: 

\begin{eqnarray}
&& A_t = \frac{B}{2} ( g_{t \phi} + 2 a g_{tt} ) - \frac{Q}{2} g_{tt} -\frac{Q}{2} \label{At}, \nonumber\\  
    && A_{\phi} = \frac{B}{2} ( g_{\phi \phi } + 2 a g_{t \phi} ) - \frac{Q}{2} g_{t \phi}
    \label{Aphi}. 
\end{eqnarray}

For a non-charged compact object, the value of $Q$ is zero. Using the above equations, we derive the equations of motion for the charged particle in the presence of an external magnetic field to study the behaviour of negative energy orbits.


\section{NEGATIVE ENERGY ORBITS and energy extraction}
\label{sec_neo}

The negative energy orbits are the orbits in which the particle's total energy is negative with respect to the asymptotic observer. For example, consider a classical system consisting of a central object and a particle. If the particle is far away from the central object, the potential energy of the particle can be considered to be zero. If the particle has zero kinetic energy near the central object then the particle's total energy can be considered negative for the asymptotic observers. If a particle gains energy equivalent to the negative energy, the particle can escape to infinity and be free from the influence of the central object. For a detailed explanation of negative energy orbits, refer \cite{Patel:2023efv}. \\

Let's delve into the fascinating Penrose process. Imagine a particle splitting into two inside the ergoregion of the compact object. One of these particles enters orbit with a negative energy state and eventually falls into the central black hole due to gravitational pull, while the other particle gets the energy equivalent to the frame-dragging energy of the particle and the rest mass energy such that the particle escapes to infinity. As previously discussed, in the case of Kerr black holes, negative energy orbit only exists inside the ergoregion. Similarly, particles can have negative energy due to electromagnetic fields around compact objects. So here, we will discuss the negative energy orbits of the charged particle around different compact objects in the presence of the non-zero electromagnetic field. \\

Consider the general metric of the rotating compact object \(g_{\mu \nu}\) surrounded by a non-zero electromagnetic four-vector field $A_\mu$.

\begin{equation}
    ds^2 =  g_{ t t }\,dt^2 + 2\,g_{t \phi }\,dt\,d\phi + g_{ rr }\,dr^2 + g_{ \theta \theta }\,d\theta^2 + g_{ \phi \phi}\, d\phi^2.
    \label{metric}
\end{equation}

 For simplicity, we assume that the particle's charge is $q$, the mass is $m$, and the particle is moving in the equatorial plane, where $q<<Q$ and $m<<M$ respectively. One can write conserved energy and angular momentum as,

\begin{equation}
    -E = g_{tt}\,\dot{t} + g_{t \phi }\,\dot{\phi} + q\,A_{t},
    \label{E}
\end{equation}

\begin{equation}
    L = g_{ \phi \phi }\,\dot{\phi} + g_{ t \phi }\,\dot{t} + q\,A_\phi.
    \label{L}
\end{equation}

From Eqs. (\ref{E}) and (\ref{L}), one can obtain expressions for the four-velocity components of the charged particle, $ u^t $ and $ u^\phi $, as:

\begin{equation}
    u^{\phi} = -\left( \frac{ E\,g_{t \phi} + L\,g_{tt}}{g^2_{t\phi} - g_{tt} \, g_{\phi \phi}} \right) - q \left( \frac{A_t \, g_{t\phi} - A_\phi \, g_{tt}}{ g^2_{t\phi} - g_{tt} \, g_{ \phi \phi } }\right),
    \label{uphi}
\end{equation}

\begin{equation}
    u^t = - \left( \frac{E \, g_{\phi \phi} + L \, g_{ t \phi }}{ g_{tt} \, g_{\phi \phi} - g^2_{ t \phi }} \right) - q \left ( \frac{ A_t \, g_{ \phi \phi } - A_{\phi} \, g_{t \phi} } { g_{tt} \, g_{ \phi \phi } - g^2_{ t \phi} } \right).
    \label{ut}
\end{equation}

\begin{figure*}[ht!]
\centering
\subfigure[Energy extraction efficiency in Kerr black hole using magnetic Penrose process. Where, $M\,=\,1, a\,=\,0.1$.]
{\includegraphics[width=7.5cm]{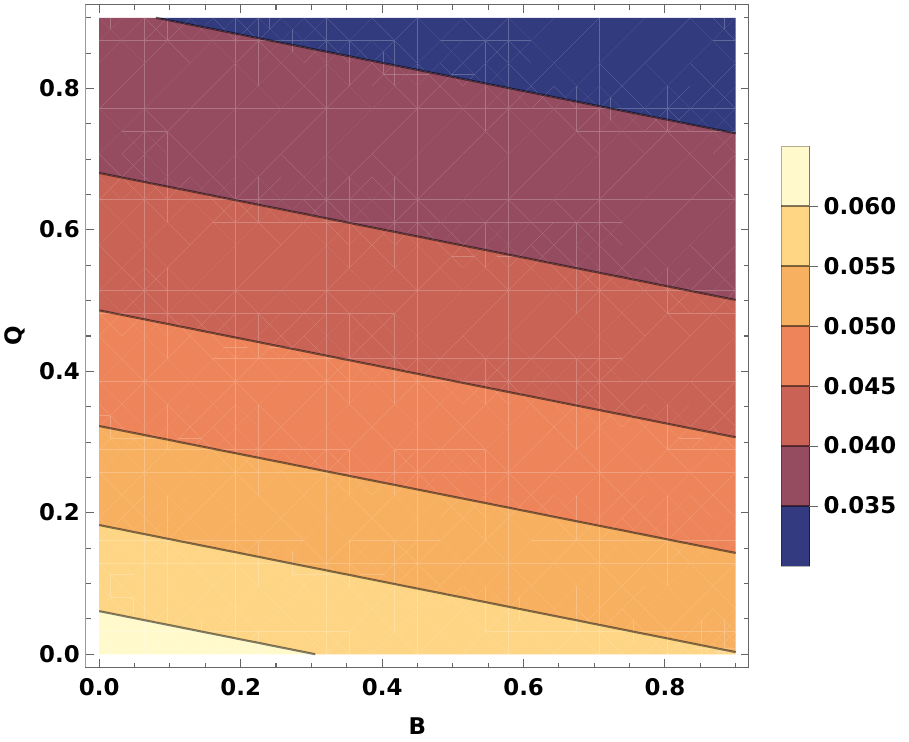}\label{fig:031}}
\hspace{1cm}
\subfigure[Energy extraction efficiency in Kerr black hole using magnetic Penrose process. Where, $M,=\,1, a\,=\,0.9$.]{\includegraphics[width=7.5cm]{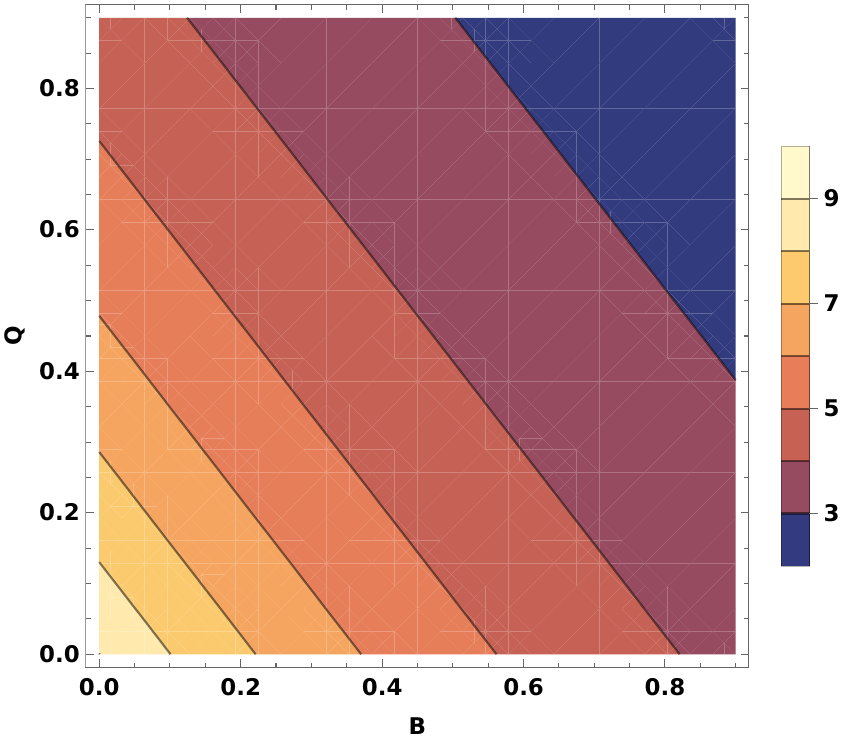}}\label{fig:03}
\subfigure[Energy extraction efficiency in Kerr black hole using magnetic Penrose process. Where, $a\,=\,0.99$.]
{\includegraphics[width=7.5cm]{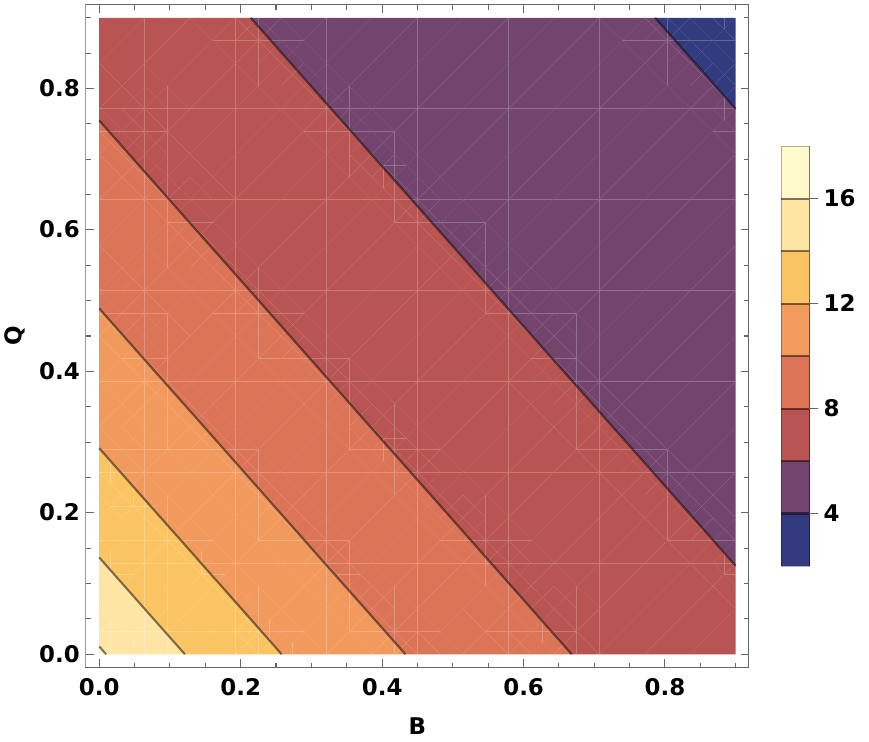}\label{fig:033}}
\hspace{1cm}
\subfigure[Energy extraction efficiency in Kerr extremal black hole using magnetic Penrose process. Where, $a\,=\,1$.]
{\includegraphics[width=7.5cm]{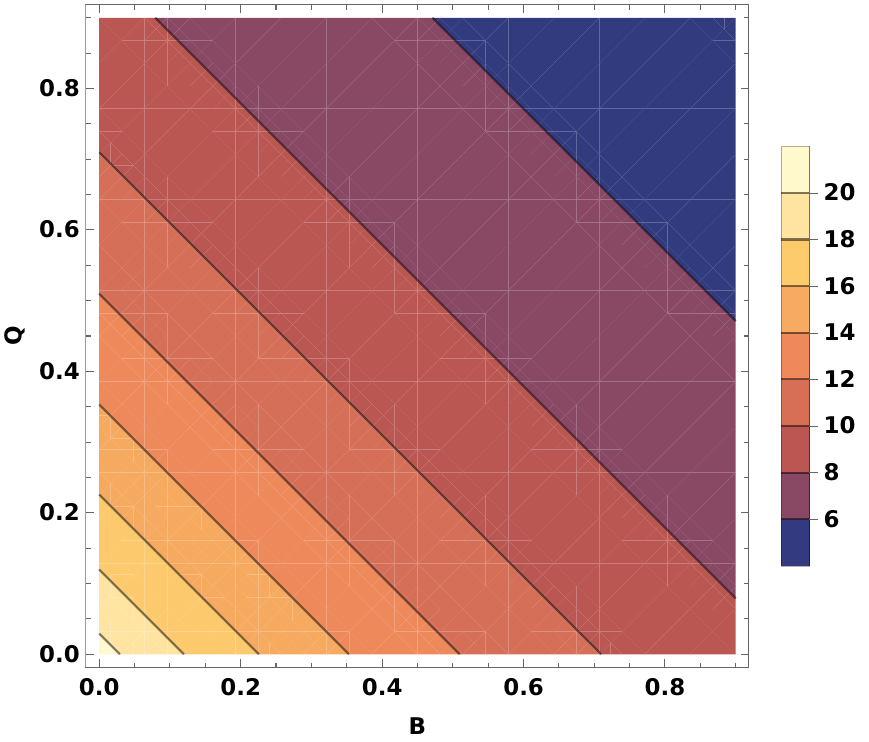}\label{fig:034}}
\caption{Representation of changes in energy extraction efficiency (bar on right side of the Figs.) with respect to the magnetic field (B) and charge (Q). }\label{fig:030}
\end{figure*}

Now consider the circular motion of the charged particle in $\theta = \pi/2$ plane, while considering $u_\alpha u^\alpha = -1 $, we can write Eq. (\ref{metric}) as,

\begin{figure*}[ht!]
\centering
\subfigure[$M\,=\,1, a\,=\,0.1$.]
{\includegraphics[width=5.7cm]{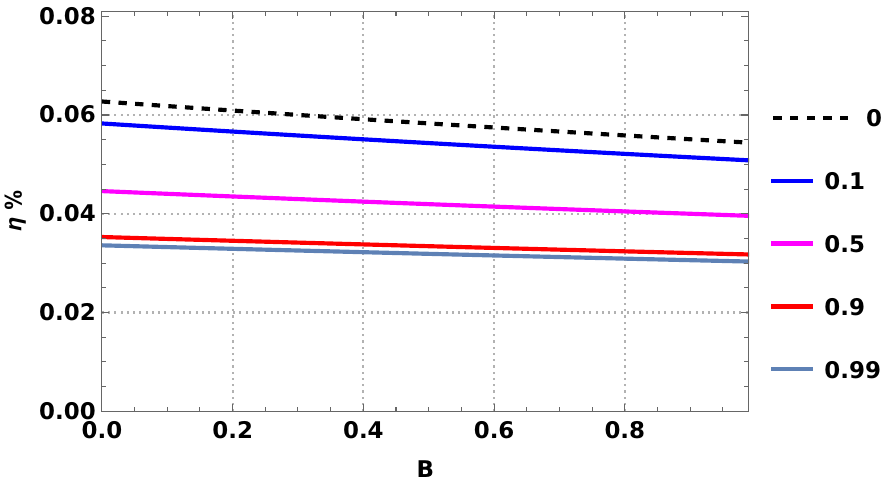}\label{fig:041}}
\subfigure[$M\,=\,1, a\,=\,0.99$.]
{\includegraphics[width=5.7cm]{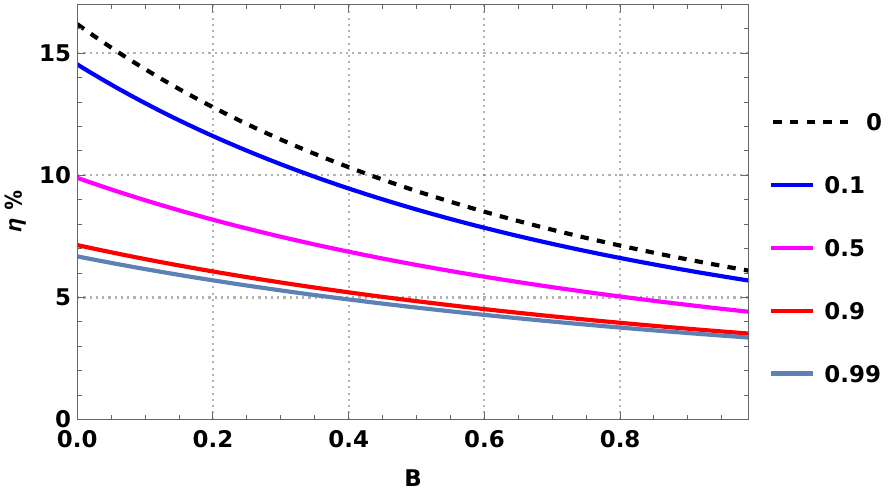}\label{fig:042}}
\subfigure[$M\,=\,1, a\,=\,1$.]
{\includegraphics[width=5.7cm]{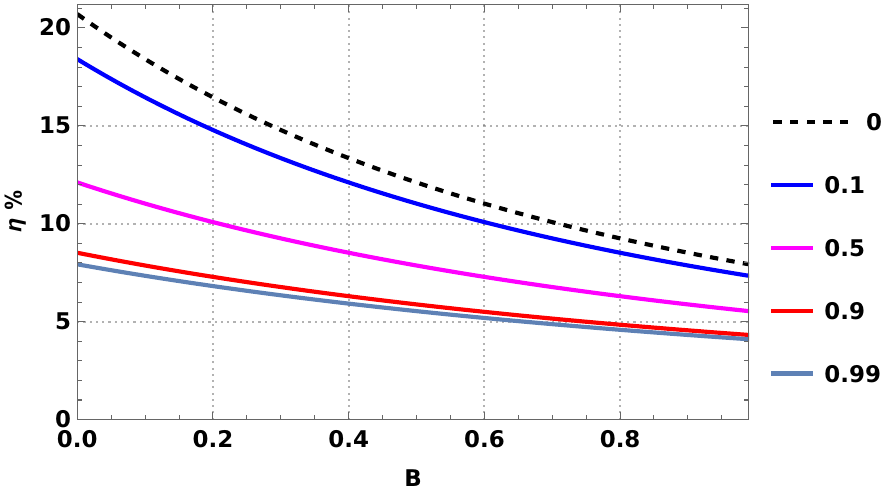}\label{fig:043}}
\caption{The change in energy extraction efficiency with respect to the represented field is shown in Figs. (bar on right side of the Figs. represents the charge (Q)). }\label{fig:040}
\end{figure*}

\begin{figure*}[ht!]
\centering
\subfigure[$a\,=\,0.3, g\,=\,0.8$.]
{\includegraphics[width=7.5cm]{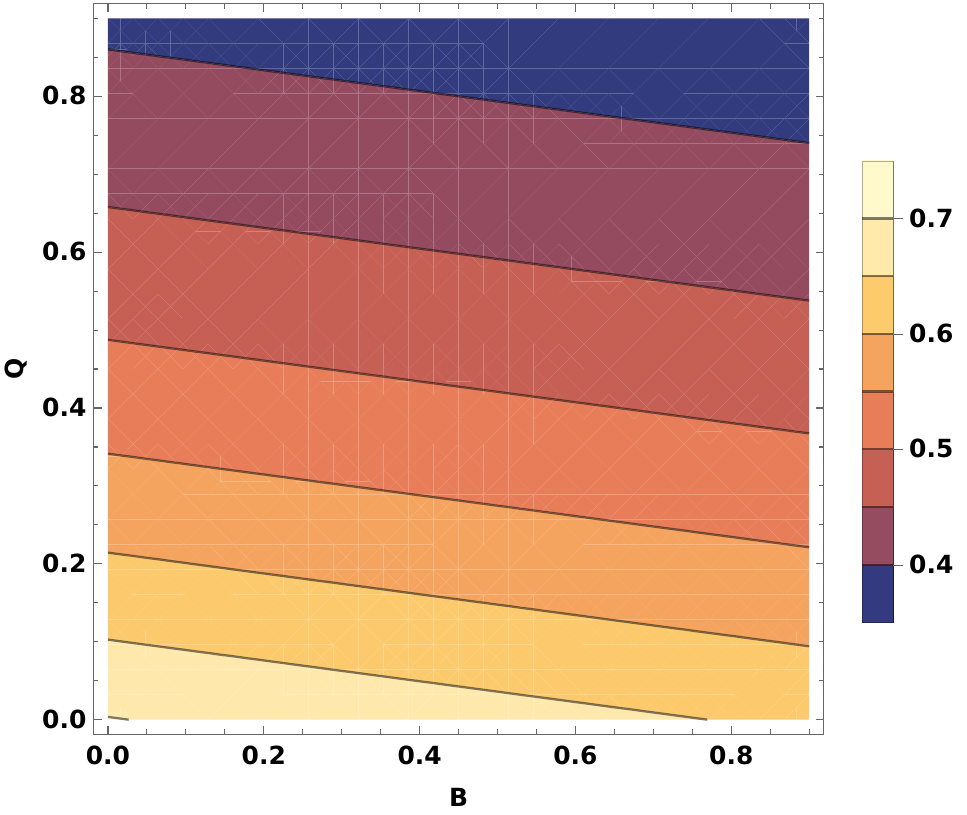}\label{fig:051}}
\hspace{1cm}
\subfigure[$a\,=\,0.7, g\,=\,0.7$.]
{\includegraphics[width=7.5cm]{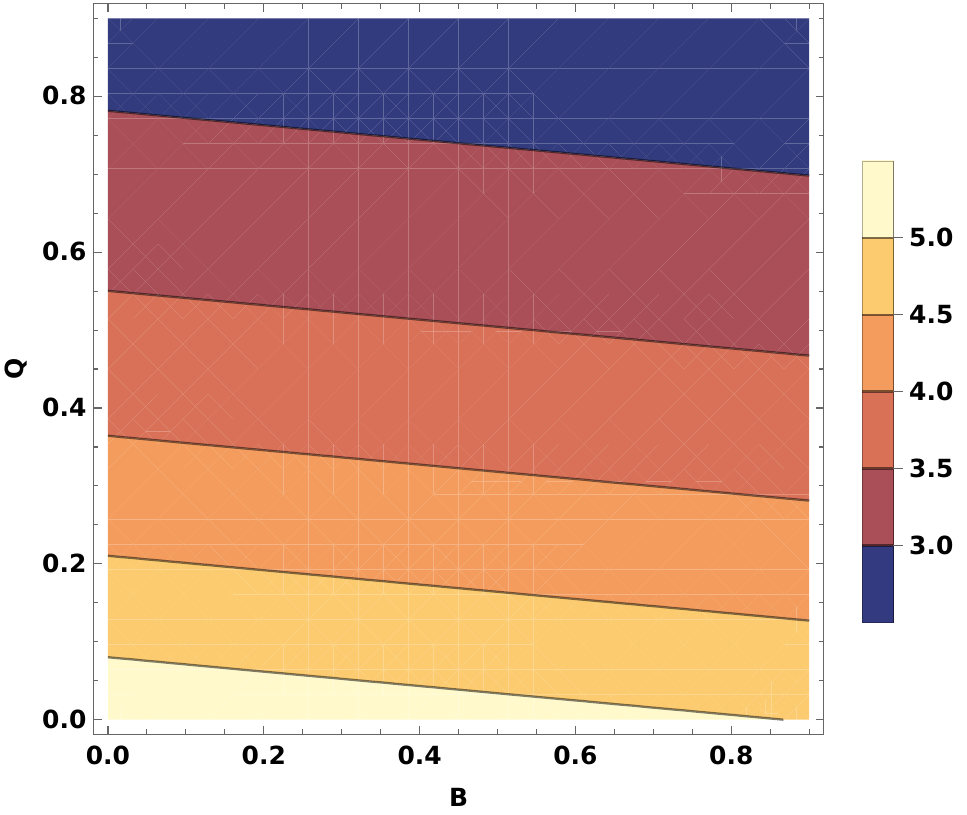}}\label{fig:05000}
\subfigure[$a\,=\,0.8, g\,=\,0.6$.]
 {\includegraphics[width=7.5cm]{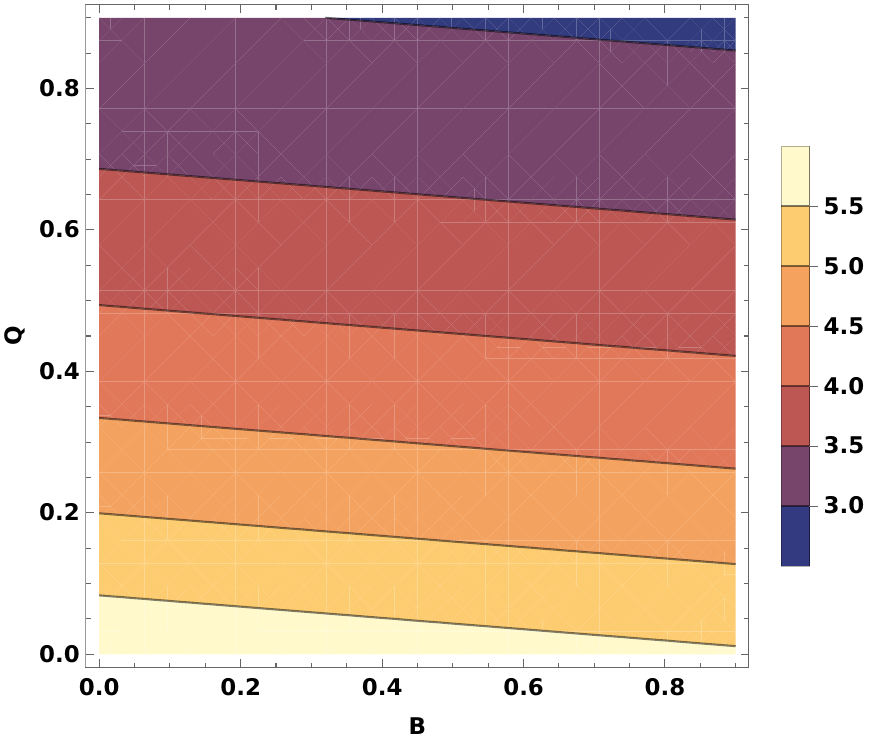}\label{fig:053}}
\hspace{1cm}
\subfigure[$a\,=\,0.8, g\,=\,0.1$.]
{\includegraphics[width=7.5cm]{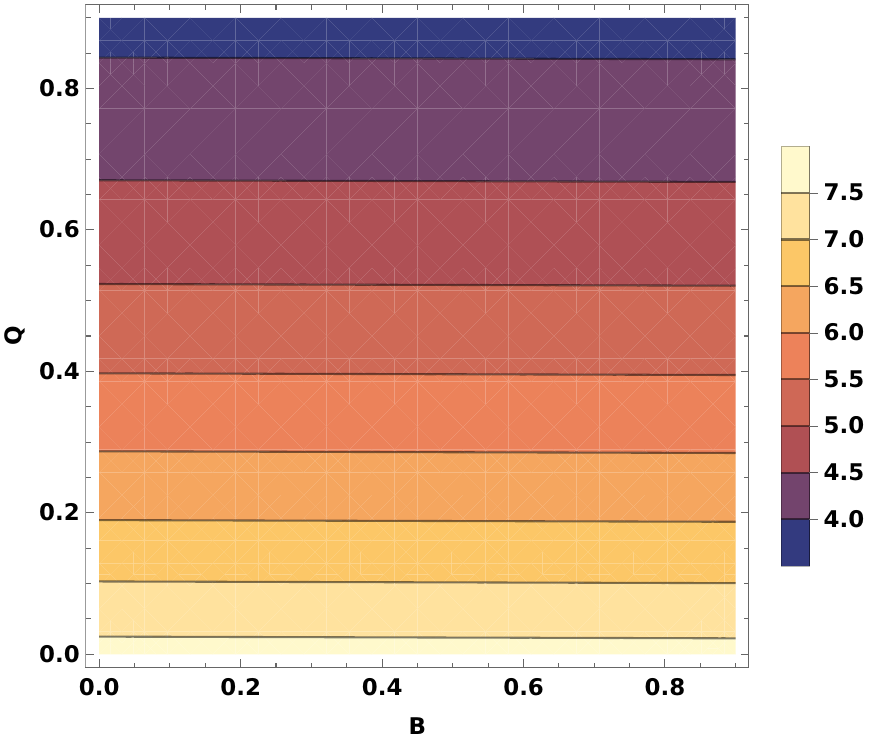}\label{fig:054}}
\subfigure[$a\,=\,0.9, g\,=\,0.4$.]
 {\includegraphics[width=7.5cm]{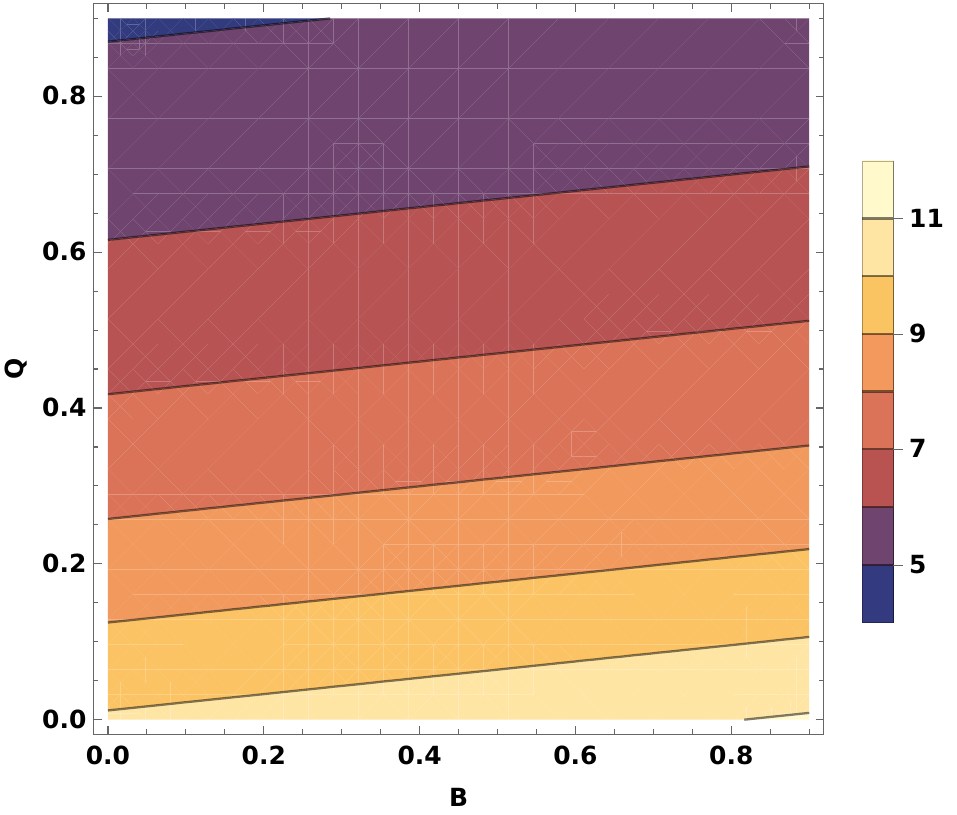}\label{fig:055}}
\hspace{1cm}
\subfigure[$a\,=\,0.99, g\,=\,0.2$.]
{\includegraphics[width=7.5cm]{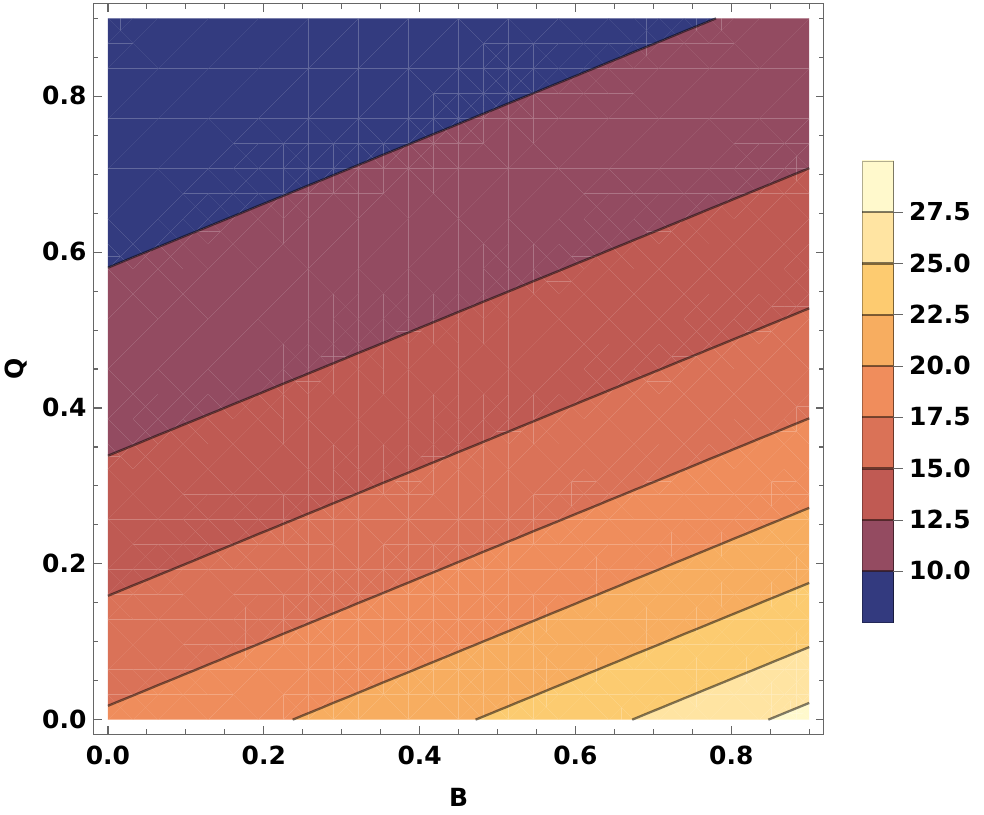}\label{fig:056}}
\caption{Energy extraction efficiency in rotating  Hayward black hole. Representation of changes in energy extraction efficiency (bar on the right side of Figs.) with respect to the magnetic field (B) and charge (Q). }\label{fig:050}
\end{figure*}

\begin{equation}
    g_{tt} \, ( u^t )^2 + 2 \, g_{ t \phi } \, (u^t) \, (u^\phi) + g_{ \phi \phi } \, (u^\phi)^2 = -1.
    \label{metric1}
\end{equation}

\begin{widetext}

\begin{table}[h!]
\centering
\begin{tabular}{| c | c | c | c | c | c | c | c | c | c ||}
    \hline
&  &  & a = 0.3 & a = 0.5  & a = 0.7 & a = 0.8 & a = 0.9  & a = 0.99 & a = 1  \\ \hline
& No.& g & - & - & - & - & - & - & - \\ \hline  \hline
& 1 & 0   & 0.5384 & 1.6109 & 3.6511 & 5.3830 & 8.2623 & 15.1030 & 19.5122   \\ \hline
& 2 & 0.2 & 0.5396 & 1.6154 & 3.6670 & 5.4170 & 8.3615 & 16.7116 & -  \\ \hline
& 3 & 0.4 & 0.5483 & 1.6485 & 3.7865 & 5.6850 & 9.2949 & - & - \\ \hline
& 4 & 0.6 & 0.5749 & 1.7535 & 4.2227 & 7.0161 & - & - & - \\ \hline
& 5 & 0.7 & 0.6013 & 1.8655 & 4.8811 & - & - & - & - \\ \hline
& 6 & 0.8 & 0.6449 & 2.0774 & - & - & - & - & - \\ \hline 
    \end{tabular}
    \caption{Energy extraction efficiency ($\eta \%$) in rotating Hayward black hole; when, B = 0.1, Q = 0.1}
    \label{tab:1}
\end{table}
 
\begin{table}[h!]
\centering
\begin{tabular}{| c | c | c | c | c | c | c | c | c | c ||}
    \hline
&  &  & a = 0.3 & a = 0.5 & a = 0.7 & a = 0.8 & a = 0.9 &  a = 0.99 & a = 1  \\ \hline
& No. & g & - & - & - & - & - & - & - \\ \hline  \hline
& 1 & 0   & 0.3059 & 0.8828 & 1.9655 & 2.9134 & 4.5950 & 9.2372 & 12.7099  \\ \hline
& 2 & 0.2 & 0.3065 & 0.8853 & 1.9742 & 2.9326 & 4.6552 & 10.4448 & -  \\ \hline
& 3 & 0.4 & 0.3115 & 0.9034 & 2.0399 & 3.0846 & 5.2282 & - & - \\ \hline
& 4 & 0.6 & 0.3266 & 0.9611 & 2.2808 & 3.8493 & - & - & - \\ \hline
& 5 & 0.7 & 0.3415 & 1.0225 & 2.6464 & - & - & - & - \\ \hline
& 6 & 0.8 & 0.3662 & 1.1388 & - & - & - & - & - \\ \hline
    \end{tabular}
    \caption{Energy extraction efficiency ($\eta \%$) in rotating Hayward black hole; when, B = 0.9, Q = 0.9}
    \label{tab:2}
\end{table}
       
\end{widetext}

Using Eqs. (\ref{ut}) and (\ref{uphi}) in Eq. (\ref{metric1}) and solving for $E$, we obtain a general expression describing the energy state of a particle orbiting a compact object given by,

\begin{equation}
 E =   \frac{-2 \, L \, g_{\text{t$\phi $}}+g_{\phi \phi } \, \left(q \, Q-X \, g_{\text{tt}}\right)+X \, g_{\text{t$\phi $}}^2}{2 \, g_{\phi
   \phi }},
\end{equation}

where,
\begin{widetext}
\begin{equation}
  X =  \sqrt{\frac{q \, \left(g_{\text{t$\phi $}} \, (2 \, a \, B-Q)+B \, g_{\phi \phi }\right) \left(q \, g_{\text{t$\phi $}} \, (2 \, a \, B-Q)+ B \, q \,   g_{\phi \phi }-4 \, L\right)+4 \, \left(g_{\phi \phi }+L^2\right)}{g_{\text{t$\phi $}}^2-g_{\text{tt}} \, g_{\phi \phi
   }}}+2 \, a \, B \, q - q \, Q.
\end{equation}
\end{widetext}

Here, we consider the negative energy orbits for $ E<0 $ for $ L<0$, and the particle's charge and charge in the geometry is such that condition $ q\, Q<0 $ is fulfilled. The value of $L$ is negative because, in the rotating compact object, two cases are considered for the motion of a particle. In the first case, a particle is orbiting around a compact object in the same direction as the spin of the compact object $( L > 0 )$. The other case is when the particle is orbiting in the opposite direction of the spin of the compact object $( L < 0 )$. Hence, a particle coming from the same direction as the object's spin will eventually move in the object's direction. Meanwhile, the particle coming from the opposite direction experiences the frame-dragging effect, causing it to eventually move in the same direction as the object's spin. The other condition with charge 
($ q\, Q < 0 $) is considered to avoid the repulsive interaction between the particle's charge and the geometry's induced charge. \\

Let us take a quick review of the magnetic Penrose process, as we need it to study the energy extraction process from rotating compact objects surrounded by a source-less magnetic field. Consider the motion of a charged particle near a rotating compact object with charge $q_1$, mass $m_1$ and energy $E_1$. Now similar to the Penrose process, the particle splits into two particles, one having mass $m_2$, charge $q_2$, energy $ E_2$ and the other particle with mass $m_3$, charge $q_3$, energy $ E_3$ such that $ q_1 > q_2,\,q_3$ and $m_1 > m_2,\,m_3$. We assume that the particle with charge $q_3$ and energy $E_3$ will enter a negative energy state $ E_3 < 0$ and fall into the compact object while the other particle with charge $q_2$ and energy $E_2$ gains this extra energy and escapes to infinity with $E_2>0$. Considering different conservation laws during the process, one can write, considering the conservation laws for energy, mass,  electric charge, linear momentum and angular momentum, we can write,  

\begin{eqnarray}
   && E_1 = E_2 + E_3, \quad  q_1 = q_2 +q_3, \quad   m_1 = m_2 + m_3,\nonumber\\
   &&  P^\mu_1 = P^\mu_2 + P^\mu_3, \qquad L_1 = L_2 + L_3.
\end{eqnarray}

One can define the angular velocity of the particle as,

\begin{equation}
    \Omega = \frac{u^\phi} {u^t} = \frac{\dot{\phi}} {\dot{t}},
\end{equation}

and using conservation $ u^\phi = \Omega \beta / \alpha $ where, 
\begin{eqnarray}
    \alpha = \epsilon + \frac{q A_t}{m}, \,\, \beta = g_{tt} + \Omega \, g_{t \phi}, 
   \,\, \epsilon = E/m.
\end{eqnarray}

One can write the energy of the escaped particle as,

\begin{equation}
    E_2 = \chi \, ( E_1 + q_1 A_t ) - q_2 \, A_t,
    \label{E2}
\end{equation}

where $\chi$ and $\beta$ are defined as,
\begin{eqnarray}
   \chi = \frac{\Omega_1 - \Omega_3}{\Omega_2 - \Omega_3} \, \frac{\beta_2}{\beta_1}, \quad  \beta_i = g_{tt} + \Omega_i \, g_{t \phi}, 
\end{eqnarray}

and $\Omega_i$ is the angular velocity of the $ i^{th}$ particle. The angular velocity of different particles can be written as,

\begin{eqnarray}
   && \Omega_1 = \frac{ -g_{ t\phi } \, ( \pi^2_\nu + g_{ t t }) + \sqrt{( \pi^2_\nu + g_{ t t } ) \, ( g^2_{ t \phi }- g_{ t t } \, g_{ \phi \phi } ) \, \pi^2_\nu} } { \pi^2_\nu \, g_{ \phi \phi }  + g^2_{ t \phi }}
    \label{omega1}, \nonumber\\
  && \Omega_2 = \frac{ -g_{ t \phi } - \sqrt{ g^2 _{t\phi} - g_{ t t } \, g_{\phi \phi } } }{g_{ \phi \phi }},
   \label{omega2}\nonumber\\
 &&\Omega_2 = \frac{ -g_{ t \phi } - \sqrt{ g^2 _{t\phi} - g_{ t t } \, g_{\phi \phi } } }{g_{ \phi \phi }},
\label{omega2}  
 \end{eqnarray}

where, $\pi_\nu = -(\epsilon + qA_t/m)$. The efficiency can be defined as,
\begin{equation}
    \eta = \frac{ E_3 - E_1}{ E_1 } = \frac{ - E_2}{ E_1 },
    \label{eta1}
\end{equation}
by using (\ref{E2}) and algebraic manipulation, one can get,

\begin{equation}
    \eta =  \frac{\chi \,  q_1  \, A_t - q_3 \, A_t}{ E_1 } + \chi - 1.
    \label{eta2}
\end{equation}
For maximum energy extraction efficiency, one can use above Eq. (\ref{eta2}) at the event horizon in the black hole case.\\

\section{Discussion}
\label{Sec_discussion}
The presented study introduces the electromagnetic field around compact objects such as rotating Simpson-Visser and Hayward black hole. A comparison is made with the Kerr and rotating Janis-Newman-Winicour naked singularity. The negative energy orbits are observed, and the magnetic Penrose process is discussed for energy extraction. We list here the salient observations from our current endeavour:

\begin{itemize}

    \item When an electromagnetic field is considered around a compact object, the region of negative energy orbits is larger compared to the negative energy orbits without an electromagnetic field. In addition, the effective ergoregion is also present (negative energy orbits exist outside the ergoregion of the compact object) in such a scenario. Here, effective ergoregion exists due to the presence of the electromagnetic field \cite{GRuffini1973}.

    \item Figs.\,(\ref{fig-1}) shows the change in the energy ($E$) with respect to the radial coordinate ($r$) for different values of regularization parameter ($l$), magnetic field ($B$) and charge ($Q$). The region of the negative energy orbits depends upon the value of the regularization parameter and the intensity of the magnetic field, charge, and spin ($a$).\, As shown in Figs.\,(\ref{fig-113}\,-\,\ref{fig-115}), as we increase the value of the regularisation parameter, the region of negative energy orbits decreases, and if we continue to increase the value of regularisation parameter, negative energy orbits will disappear.\, If we continue to increase the value of the magnetic field and charge, it causes an increment in the region of negative energy orbits. 

    \item The energy extraction efficiency using the magnetic Penrose process in the Kerr and Simpson-Visser geometries surrounded by electromagnetic fields are represented in Fig.\,(\ref{fig:030}),\,Fig.\,(\ref{fig:034}) represent the case of extremal black hole. For rotating Simpson-Visser black hole, we have shown the change in the efficiency with respect to the magnetic field for different values of the spin parameter Fig.\,(\ref{fig:040}).   
    
    \item In the Penrose process, the maximum efficiency for a Kerr black hole is $20.71\%$. From Figs.\,(\ref{fig:030}) and (\ref{fig:040}), one can observe that as the magnetic field and charge increases, the efficiency decreases. Similar observations are shown in \cite{Chakraborty:2024aug}, where a particular type of magnetic field profile causes a reduction in energy extraction efficiency.
    
    \item We find in our present study that the magnetic Penrose process does not depend on the regularisation parameter of the Simpson-Visser geometry; similar conclusions were made in \cite{Patel:2022jbk,Franzin:2022iai} for the case of superradiance and Penrose process. Which shows that extracted energy efficiency is the same in both Kerr and Simpson-Visser geometry (Fig.\,(\ref{fig:030})). Thus,  the energy extraction efficiency of the Simpson-Visser geometry remains indistinguishable from the Kerr black hole. However, ergoregion and negative energy orbits are different in the rotating Simpson-Visser black hole (depending on the regularisation and spin parameter) from the Kerr black hole (depending on the spin parameter).    
    
    \item The energy extraction efficiency from the rotating Hayward black hole is presented in Tab: (\ref{tab:1}) and (\ref{tab:2}) with different spin and deviation parameters for particular magnetic fields and charges. The dashed portion for particular deviation and spin parameters indicates the absence of an event horizon. However, in the Penrose and magnetic Penrose process, a rotating Hayward black hole has maximum energy extraction efficiency compared to the Kerr black hole case \cite{Khan:2021wzm}. A noticeable thing is that energy extraction efficiency is highest in the Kerr black hole when the magnetic field and charge are lowest Fig.\,(\ref{fig:030}). On the other hand, in the rotating Hayward black hole, the relation between the magnetic field and charge changes as the deviation and spin parameter changes Fig.\,(\ref{fig:050}). In Figs.\,(\ref{fig:051}\, and\,(\ref{fig:05000})), the energy extraction efficiency is highest when the magnetic field and charge are lowest. The change in energy extraction efficiency with different deviation parameters and at particular spin parameters ($a\,=\,0.8$) is shown in Figs.\,((\ref{fig:053}) and (\ref{fig:054})). In Fig.\,(\ref{fig:054}), we can see that with the lowest deviation parameter, the change in the magnetic field does not affect the energy extraction efficiency. Nevertheless, in Figs.\,((\ref{fig:055}), (\ref{fig:056})), one can observe that the energy extraction efficiency is maximum when the magnetic field is maximum and the charge is minimum. 
    
    \item Our study shows that rotating Hayward black holes achieve higher energy extraction efficiency than rotating Simpson-Visser and Kerr black holes in both the Penrose and magnetic Penrose processes. In comparing different types of Kerr black hole mimickers, which include regular black holes (Simpson-Visser, Hayward) and naked singularities (Kerr, Janis-Newman-Winicour), it is found that the highest energy extraction efficiency is achieved in the case of a Kerr naked singularity ($a > M$)\, \cite{Stuchlik1980}. Following that is the case of a rotating Janis-Newman-Winicour naked singularity \cite{Patel:2023efv}. Subsequently, the rotating Hayward black hole demonstrates higher efficiency than the Kerr black hole. In summary, the order of energy extraction efficiency from highest to lowest is as follows: Kerr naked singularity, rotating Janis-Newman-Winicour naked singularity, rotating Hayward black hole, rotating Simpson-Visser black hole and finally, the Kerr black hole.
\end{itemize}

\section{ACKNOWLEDGMENTS}
VP would like to acknowledge the support of the
Council of Scientific and Industrial Research (CSIR, India, Ref: 09/1294(18267)/2024-EMR-I) for funding the
work.

\end{document}